\documentclass{jfm}

\usepackage{cite}
\usepackage{url}

\usepackage{graphicx}
\usepackage{natbib}
\usepackage{amssymb,amsmath}

\newcommand{\C}{\mathbb C}
\newcommand{\e}{\eqref}

\ifCUPmtlplainloaded \else
  \checkfont{eurm10}
  \iffontfound
    \IfFileExists{upmath.sty}
      {\typeout{Found AMS Euler Roman fonts on the system,
                   using the 'upmath' package.}%
       \usepackage{upmath}}
      {\typeout{Found AMS Euler Roman fonts on the system, but you
                   dont seem to have the}%
       \typeout{'upmath' package installed. JFM.cls can take advantage
                 of these fonts, if you use 'upmath' package.}%
       \providecommand\upi{\upi}%
      }
  \else
    \providecommand\upi{\upi}%
  \fi
\fi

\ifCUPmtlplainloaded \else
  \checkfont{msam10}
  \iffontfound
    \IfFileExists{amssymb.sty}
      {\typeout{Found AMS Symbol fonts on the system, using the
                'amssymb' package.}%
       \usepackage{amssymb}%
       \let\le=\leqslant  
       \let\ge=\geqslant  
      }{}
  \fi
\fi

\ifCUPmtlplainloaded \else
  \IfFileExists{amsbsy.sty}
    {\typeout{Found the 'amsbsy' package on the system, using it.}%
     \usepackage{amsbsy}}
    {\providecommand\boldsymbol[1]{\mbox{\boldmath $##1$}}}
\fi

\newcommand\I{\mbox{i}}
\newcommand\D{\mbox{d}}
\newsavebox{\astrutbox}
\sbox{\astrutbox}{\rule[-5pt]{0pt}{20pt}}

\title[Branch cuts of Stokes wave on deep water. Part II
]{Branch cuts of Stokes wave on deep water. Part II: Structure and location of branch points in infinite set of sheets of Riemann
surface }

\author[P.\,M.~Lushnikov]
{Pavel M. Lushnikov
\thanks{Email address for correspondence:
plushnik@math.unm.edu}}

\affiliation{
Department of Mathematics and Statistics, University of New Mexico,
Albuquerque, MSC01 1115, NM, 87131, USA }

\begin{document}

\maketitle

\begin{abstract}
Stokes wave  is  a finite amplitude periodic gravity wave propagating with constant velocity in inviscid fluid.
 Complex analytical structure of Stokes wave is analyzed using a conformal mapping of a free fluid surface  of Stokes
 wave into the real line with fluid domain mapped into the lower complex half-plane.
 There is  one square root branch point per   spatial period of Stokes located in the upper complex half-plane at the distance $v_c$ from the real axis.  The increase of Stokes wave height results in approaching $v_c$ to zero with the limiting Stokes wave formation at $v_c=0.$   The limiting Stokes wave has $2/3$ power law singularity forming $2\upi/3$ radians angle on the
crest  which is qualitatively different from the square root
singularity valid for arbitrary small but nonzero $v_c$ making the
limit of zero $v_c$ highly nontrivial. That limit is addressed by
crossing a branch cut of a square root into the second and subsequently higher
sheets of Riemann surface to find coupled square root singularities
at the distances  $\pm v_c$ from the real axis at each sheet. The
number of sheets is infinite and   the analytical continuation of
Stokes wave into all these sheets is found together with the series
expansion in half-integer powers  at singular points
within each sheet. It is  conjectured that non-limiting Stokes wave
at the leading order  consists of the infinite number of nested
square root singularities which also implies the existence in the
third and higher sheets of the additional square root singularities
away from the real and imaginary axes. These nested square roots
form $2/3$ power law singularity of the limiting Stokes wave as
$v_c$ vanishes.

\end{abstract}


\section{Introduction}\label{sec:introduction}

\begin{figure}
\includegraphics[width=1.02\textwidth]{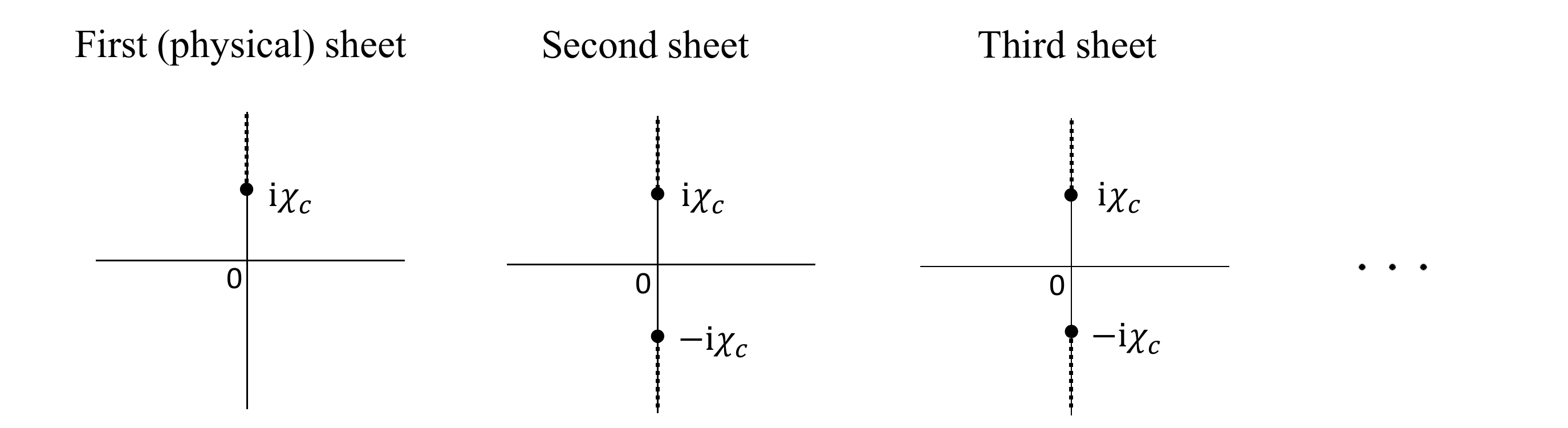}
\vspace{-0.1cm}
 \caption{  A schematic of Riemann surface sheets for non-limiting Stokes
 wave in the complex variable $\zeta$  \e{zetadef} near the origin. The first (physical) sheet has a single square root singularity at $\zeta=\I\chi_c$  in the upper
 complex half-plane $\mathbb{C}^+$ with the the lower
 complex half-plane $\mathbb{C}^-$ corresponding to the domain occupied by the fluid. Other (non-physical) sheets have square root singularities at $\zeta=\pm \I\chi_c$.  Dashed lines show branch cuts. In
addition there are the singularities at $\zeta=\pm\I$ in all sheets which corresponds to $w=\infty$.  As well as starting from
the third sheet there are square root singularities away from both  real and imaginary axes at the distances  more that several
times exceeding $\chi_c$, i.e. well beyond the disks of convergence $|\zeta\pm \I \chi_c|<2\chi_c$.}
 \label{fig:sheets_sketch}
\end{figure}

In Part I
\citep{DyachenkoLushnikovKorotkevichPartIStudApplMath2016}, we
obtained Stokes wave solution numerically with high precision and
analyzed that solution using Pad\'e approximation. We showed a
convergence of Pad\'e   approximation of Stokes wave to a single
branch cut per spatial period in the upper complex half plane
$\mathbb{C}^+$ of the axillary complex variable $w$.   In this paper
we formulate the nonlinear integral equation for the jump of Stokes
wave at the  branch cut in the physical (first) sheet of Riemann
surface. We show that the Riemann surface of Stokes has infinite
number of sheets as sketched in Figure \ref{fig:sheets_sketch} and
study the structure of singularities in these sheets.

Stokes wave  is  the fully  nonlinear periodic gravity wave
propagating with the constant velocity $c$
~\citep{Stokes1847,Stokes1880}. It corresponds to two-dimensional
potential flow of an ideal incompressible fluid with free surface. Following Part I
\citep{DyachenkoLushnikovKorotkevichPartIStudApplMath2016}, we use
scaled units  at which $c=1$ for the linear gravity waves and the
spatial period is $\lambda=2\upi$.  Nonlinearity of Stokes wave
increases with the increase of $H/\lambda,$ where $H$ is the Stokes
wave height which is defined as the vertical distance from the crest
to the trough of Stokes wave.  Stokes wave has  $c>1$ and the limit  $H\to 0,
\ c\to 1$ corresponds to the  linear gravity wave.  The
Stokes wave of the greatest height  $H=H_{max}$ (also called by the
limiting Stokes wave) has the singularity in the form  of the sharp
angle of $2\upi/3$ radians on the crest~\citep{Stokes1880sup}. We
assume that singularity of the limiting Stokes wave  touches the fluid surface at $w=0$
and corresponds the following
expansion
\begin{equation} \label{w2p3}
z(w)=\I \frac{c^2}{2}-\I\left (\frac{3c}{2}\right )^{2/3}(\I w)^{2/3}+\text{h.o.t.}
\end{equation}
which ensures  the sharp angle of $2\upi/3$ radians on the
crest. Equation \e{w2p3} recovers the result of  ~\citet{Stokes1880sup}. Here  \text{h.o.t.} means higher order terms which approaches $0$ faster than $w^{2/3}$ as $w\to 0$. Also
\begin{equation} \label{zwdef}
z(w)=x(w)+\I y(w)
\end{equation}
is the conformal transformation which maps a half-strip $-\pi     \le u\le \pi$, $-\infty<v\le 0$
of the conformal variable
\begin{align} \label{wdef}
w=u+\I v
\end{align}
into a fluid domain of infinite depth $-\infty<y\le\eta(x), \
-\pi\le x\le \pi$  of the complex plane $z$ (see Figure 1 of Part I
\citep{DyachenkoLushnikovKorotkevichPartIStudApplMath2016}). Here
$x$ and $y$ are the horizontal and vertical physical coordinates,
respectively.  $y=\eta(x)$ is the surface elevation in the reference
frame moving with the speed $c$. As discussed in details in Part I,
choosing
\begin{equation}\label{xtildedef}
\begin{split}
& z(w)=w+\tilde z(w),
\end{split}
\end{equation}
with $x(w)=u+\tilde x(w)$ and  $ \tilde y(w)=v+y(w)$, ensures that $\tilde z(w)$ is  $2\pi$-periodic function
\begin{align} \label{xtildeperiodic}
\tilde z(w+2\pi)=\tilde z(w), \quad \tilde x\left (\pm\pi\right )=0.
\end{align}

It was found by \citet{MalcolmGrantJFM1973LimitingStokes} that the corner singularity \e{w2p3} might not be a simple algebraic branch point because next order term in the expansion  \e{w2p3} might be a power of the transcendental number. Rigorous results on the asymptotics near the crest of the
 limiting wave were found in Refs. \citet{AmickFraenkelTransAmerMathSoc1987,McLeodMcLeodTransAmerMathSoc1987}. These results were used
 in Refs. \citet{FraenkelArchRationMechAnal2007,FraenkelHarwinEuropeanJApplMath2010,FraenkelEuropeanJApplMath2010} to construct the exact bounds on the limiting Stokes wave
 and prove the local uniqueness using Banach''s contraction mapping
 principle. More exact bounds were provided in Ref.
 \citet{Tanveerarxiv2013}.
 However, the question if log terms in the asymptotic expansion are possible in addition to the transcendental power asked in Ref.  \citet{AmickFraenkelTransAmerMathSoc1987} remains open. The existence of  limiting Stokes wave with the
jump of the slope at the crest in $2\upi/3$ radians   was independently proven by \citet{Plotnikov1982} and \citet{AmickFraenkelTolandActaMath1982}.

 In this paper we focus on analyzing  singularities of near-limiting Stokes wave. \citet{MalcolmGrantJFM1973LimitingStokes}  showed that assuming that singularity is a power law branch point, then that singularity has to have a square root form to the leading order. \citet{TanveerProcRoySoc1991} provided much stronger result proving that the only possible singularity in the finite complex upper half-plane is of square root type. Ref. \citet{PlotnikovToland2002} discusses the existence of
a unique square root singularity above crests. The existence of only
one square root singularity per period in a finite physical complex
plane   was also confirmed in Ref. \citep{{DLK2013}} and Part I
\citep{DyachenkoLushnikovKorotkevichPartIStudApplMath2016}   by
analyzing the numerical solution for Stokes wave.

We now consider an additional conformal transformation between the complex plane   $w=u+\I v$  and  the complex plane for the new variable
\begin{equation} \label{zetadef}
\zeta=\tan\left (\frac{w}{2}\right )
\end{equation}
which maps the strip $-\upi <Re(w)<\upi$ into  the complex $\zeta$
plane. In particular, the line segment $-\pi<w<\pi$ of the real line
$w=u$   maps into the entire real line $(-\infty,\infty)$ in the
complex $\zeta$-plane as shown in  Figure~5 of Part I
\citep{DyachenkoLushnikovKorotkevichPartIStudApplMath2016}. Vertical
half-lines $w=\pm\pi+\I v, \ 0<v<\infty$ are mapped into a branch
cut   $\I<\zeta<\I\infty.$ In a similar way, vertical half-lines
$w=\pm\pi+\I v, \ -\infty<v<0$ are mapped into a branch cut
$-\I\infty<\zeta<-\I.$ However, $2\pi-$periodicity of $\tilde z(w)$
\e{xtildeperiodic} allows to ignore these two branch cuts  because
$\tilde z(w)$ is continuous across them.  Complex infinities $w=\pm
\I\infty $ are mapped into $\zeta=\pm \I$.  An unbounded interval
$[\I v_c,\I\infty),$ $v_c>0$ is mapped into a finite interval
$[\I\chi_c,\I)$ with
\begin{align} \label{chicdef}
\chi_c=\tanh{\frac{ v_c}{2}}.
\end{align}
The transformation \e{zetadef} takes care of $2\pi-$periodicity of Stokes wave so that the function $z(\zeta)$ defined in the complex plane $\zeta\in \mathbb{C}$ corresponds to  the function $z(w)$ defined in the strip $-\pi     < Re(w)=u< \pi.$ Here and below we abuse notation and use the same symbol $z$ for both functions of $\zeta$ and $w$ (and similar for other symbols).
The additional advantage of using the mapping  \e{zetadef} is the compactness of the interval $(\I\chi_c,\I)$ as mapped from the infinite interval $(\I v_c,\I\infty)$.
 Note that the mapping  \e{zetadef} is different from the commonly used (see e.g. \cite{SchwartzJFM1974,Williams1981,TanveerProcRoySoc1991}) mapping $\zeta =\exp{(-\I w)}$ (maps the strip  $-\upi \le Re(w)<\upi$   into the unit circle). That exponential map   leaves the interval $(\I v_c,\I\infty)$ infinite in $\zeta$ plane.

The main result of this paper  is that it was found an infinite number of sheets of  Riemann surface with square root branch points located at $\zeta=\pm \I \chi_c$ starting from the second sheet (the first sheet has the singularity only at $\zeta=\I\chi_c$).
At each  sheet (except the first one) these singularities are coupled through complex conjugated terms which appear in the equation for Stokes wave.
In contrast, the only singularity at $\zeta=\I\chi_c$ of the first sheet (besides the singularity at $\zeta=\I$) does not have a complex conjugated sister
at  $\zeta=-\I\chi_c$  which makes that (physical) sheet distinct from all others.
It is conjectured that the leading order form of non-limiting Stokes wave has the form of the infinite number of nested square root singularities. These nested square roots  form $2/3$ power law singularity of the limiting Stokes wave as $\chi_c\to 0$.

The paper is organized as follows. In Section \ref{sec:ClosedIntegral} a closed nonlinear integral equation for Stokes wave in
terms of the density (jump) at the branch cut is derived and the numerical method to solve that integral equation is given.
Section \ref{sec:AlternativeStokeswave}  provides {an alternative form for the equation of  Stokes wave}.
 Section
\ref{sec:AsymptoticStokeswave} uses that alternative form to find an asymptotic of both Stokes wave at $Im(w)\to +\infty$ and the  jump at the branch cut.    Section \ref{sec:Numericalstructuresheets} discusses a numerical procedure to analyze the structure of sheets of Riemann surface for Stokes wave by the integration of the corresponding nonlinear ordinary differential equation (ODE) in the complex plane.  Section \ref{sec:SeriesexpansionsStructureRiemannsurface} derives the analytical expressions for coupled series expansions at $\zeta=\pm \I\chi_c$ to reveal the structure of Riemann surface for Stokes wave. Section \ref{sec:squarerootsingularities}
analyzes possible singularities of Stokes in all sheets of Riemann surface and concludes that the only possible singularity for finite value of $w$ is the square root branch point.  Section \ref{sec:conjecturetwothirds} provides a conjecture  on recovering of $2/3$ power law of limiting Stokes wave from an infinite number of nested square root singularities of non-limiting Stokes wave in the limit $\chi_c\to 0$.
In Section \ref{sec:Conclusion} the main results of the paper are
discussed.
Appendix \ref{sec:Stokeswaveequivalnce} shows the equivalence of two forms of  equation  for Stokes wave used in the main text.  Appendix \ref{sec:restmovingframe} relates different forms of equation for Stokes wave in the rest frame and in the moving frame.
Appendix \ref{sec:TablesStokesWaveschic} provides  tables of the numerical parameters of Stokes wave.

\section{Closed integral equation for Stokes wave through the density at the branch cut}
\label{sec:ClosedIntegral}

The equation for Stokes wave was derived in  Ref. \citet{ZakharovDyachenkoPhysD1996}   and Part I
\citep{DyachenkoLushnikovKorotkevichPartIStudApplMath2016}  from
Euler's equations for  the potential flow of ideal fluid with free
surface (see also Appendices  \ref{sec:Stokeswaveequivalnce} and
\ref{sec:restmovingframe}). That equation is defined  at the real
line $w=u$  and takes the following form
\begin{equation}
\label{stokes_wave} -c^2y_u + yy_u + \hat H[y(1+\tilde x_u)] = 0,
\end{equation}
where
\begin{equation} \label{Hilbertdef}
\hat H f(u)=\frac{1}{\upi} \text{p.v.}
\int^{+\infty}_{-\infty}\frac{f(u')}{u'-u}\D u'
\end{equation}
is the Hilbert
transform with $\text{p.v.}$ meaning a Cauchy principal value of integral and subscripts in $t$ and $u$ mean partial derivatives here and further.
The Hilbert operator $\hat H$ transforms
into the multiplication operator
\begin{equation} \label{Hfk}
 (\hat H f)_k=\I\,
\text{sign}{\,(k)}\,f_k,
\end{equation}
for the Fourier coefficients (harmonics) $f_k$,
\begin{align} \label{ffourier}
f_k=\frac{1}{2\upi}\int\limits_{-\upi}^{\upi} f(u)\exp\left (-\I
ku\right )\D  u,
\end{align}
 of  the periodic function $f(u)=f(u+2\upi)$ represented through the Fourier series
\begin{equation} \label{fkseries}
f(u)=\sum\limits_{k=-\infty}^{\infty} f_k\exp\left (\I
ku\right ).
\end{equation}
Here $\text{sign}(k)=-1,0,1$ for $k<0, \ k=0$ and $k>0$, respectively.
Equation \e{fkseries} implies that
\begin{equation} \label{H2def0}
\hat H^2f=-(f-f_0),
\end{equation}
where $f_0$ is the zeroth Fourier harmonic of $f$.

It is  convenient to decompose the Fourier series \eqref{fkseries} as follows
\begin{equation} \label{fpm0}
f(u)=f^+(u)+f^-(u)+f_0,
\end{equation}
where
\begin{equation} \label{fplus}
f^+(w)=\sum\limits_{k=1}^{\infty} f_k\exp\left (\I
kw\right )
\end{equation}
is the analytical function  in $\mathbb{C}^+$ and
\begin{equation} \label{fminus}
f^-(w)=\sum\limits_{k=-\infty}^{-1} f_k\exp\left (\I
kw\right )
\end{equation}
is the analytical function  in the lower complex half-plane
$\mathbb{C}^-$. Then equation \e{Hfk} implies that
\begin{equation} \label{Hfpm}
\hat H f=\I(f^+-f^-).
\end{equation}
Also using  equation \e{Hfk} we define the operator,
\begin{equation} \label{Projectordef}
\hat P=\frac{1}{2}(1+\I \hat H),
\end{equation}
 projecting any $2\upi$-periodic function $f$ into a
function which has analytical continuation from the real line $w=u$
into $\C^-$
as follows

\begin{equation} \label{fprojected}
 \hat Pf=f^-+\frac{f_0}{2}.
\end{equation}

 One can  apply $\hat H$ to \eqref{stokes_wave} to obtain the following  closed expression for $y$, %
\begin{equation}\label{stokes_wave2}
\begin{split}
&   \left(  {c^2}\hat k - 1 \right) y -  \left( \frac{\hat k y^2}{2} + y\hat k y \right) = 0,
\end{split}
\end{equation}
where    $\hat k \equiv -\partial_u \hat H =
\sqrt{-\nabla^2}$ and we used the following relations \begin{align} \label{xytransform_der}
y_u=\hat H\tilde x_u \quad  \text{and} \quad  \tilde    x_u=-\hat Hy_u,
\end{align}
which are valid for the analytic function $\tilde z_u(w)$ satisfying
the decaying condition  $\tilde z_u(w) \to 0$ as $v\to-\infty.$ We
also assume in deriving   equation \e{stokes_wave2} from equation
\e{stokes_wave}
 that%
\begin{equation} \label{yxucondition}
\int\limits^{\upi}_{-\upi}\eta(x)\D
x=\int\limits^{\upi}_{-\upi} y(u)x_u(u) \D u=0,
\end{equation}
meaning that the mean elevation of  the free surface
is set to zero. Equation \e{yxucondition}  reflects a conservation of the total mass of fluid.
Equation  \e{stokes_wave2} was  derived in Ref. \citet{BabenkoSovietMathDoklady1987} and later was independently obtained from results of Ref. \citet{DKSZ1996} in  \citet{DLK2013}.
See also Ref. \citet{ZakharovDyachenkoPhysD1996} for somewhat similar equation.
Ref. \citet{BabenkoSovietMathDoklady1987} and subsequent developments in Refs. \citet{BuffoniDancerTolandArchRationMechAnal2000,BuffoniTolandCRAcadSciParisSrIMath2001,PlotnikovIzvAkadNaukSSSRSerMat1991,ShargorodskyTolandMemAmerMathSoc2008}
used equation of the type \e{stokes_wave2} for the analysis of bifurcations.

Equation  \e{stokes_wave2} is convenient for numerical simulation of
Stokes wave because it depends on $y$ only as detailed in Part I
\citep{DyachenkoLushnikovKorotkevichPartIStudApplMath2016}. The
operator $\hat k$ is the multiplication operator in Fourier domain
which is straightforward to evaluate numerically using Fast Fourier
Transform.

In this paper    it is however more convenient for analytical study to rewrite equation for Stokes wave in terms of the complex variable $\tilde z$. For that we apply the projector operator $\hat P$  \e{Projectordef} to equation  \e{stokes_wave}
  which  results in
\begin{equation}
\label{stokes2a}
 {c^2}{\tilde z}_u =  -\I \hat P \left [({\tilde z}-\bar{{\tilde z}})   (1+{\tilde z}_u)\right. ],
\end{equation}
where $\bar f(u)\equiv\bar f$ means complex conjugation of  the function $f(u)$.
Note that the complex conjugation $\bar f(w)$ of $f(w)$  in this paper is understood as applied with the assumption that $f(w)$ is the complex-valued function of the real argument $w$ even if $w$ takes the complex values so that
\begin{equation} \label{bardef}
 \bar f(w)\equiv \overline {f(\bar{w})}.
\end{equation}
That definition ensures
  the analytical continuation of $f(w)$ from
the real axis  $w=u$ into the complex plane of $w\in\mathbb{C}$ and similar for functions of $\zeta\in\mathbb{C}$.
If the function $f(w)$ is analytic in $\C^-$ then    $\bar{f}({w})$
 is analytic in $\C^+$ as also follows from equations \e{fpm0}-\e{fminus}.

A numerical convergence of Pad\'e approximation to the continuous density $\rho(\chi)$ of the branch cut
 was shown in Part I \citep{DyachenkoLushnikovKorotkevichPartIStudApplMath2016} together with the  parametrization of the  branch cut of Stokes wave as follows
\begin{equation}
\label{branchcutdensity} \tilde z(\zeta) = \I y_b +
\int\limits_{\chi_c}^{1} \dfrac{\rho(\chi')d\chi'}{\zeta - \I \chi'}
,
\end{equation}
where $y_b\equiv y(u)|_{u=\pm\upi}\in \mathbb{R}$ is the minimum
height of Stokes wave as a function of $x$ (or in the similar way as
the function of $u$). The density
 $\rho(\chi)$ is related to the jump $\Delta_{jump}$  of $\tilde z(\zeta)$  for crossing the branch cut at $\zeta=\I \chi$ in
counterclockwise direction as follows

\begin{equation} \label{jumprho1}
\tilde \Delta_{jump}\equiv z(\zeta)|_{\zeta=\I\chi-0}-\tilde z(\zeta)|_{\zeta=\I\chi+0}=-2\pi\rho(\chi),
\end{equation}see also Part I \citep{DyachenkoLushnikovKorotkevichPartIStudApplMath2016} for more
details on that. We now use the parametrization \e{branchcutdensity}
to study the Stokes wave equation \e{stokes2a}. We eliminate the
constant $\I y_b$   at $\zeta=\infty$  in \e{branchcutdensity} by
introducing a new function
\begin{align} \label{fdef}
f(u) = \tilde z(u)-\I y_b= \int\limits_{\chi_c}^{1}
\dfrac{\rho(\chi')d\chi'}{\zeta - \I \chi'}.
\end{align}
together with the complex conjugate
\begin{align} \label{fdefbar}
\bar f(u) =  \int\limits_{\chi_c}^{1}
\dfrac{\rho(\chi')d\chi'}{\zeta + \I \chi'}
\end{align}
which was evaluated using the definition \e{bardef}.

 Equation (\ref{stokes2a}) in the new valuable \e{fdef} takes
the following form

\begin{equation}
\label{eqn:stokes22} -\I {c^2}{f}_u   +{ \I y_b+2\I y_b} {f_u+{ f}
{f_u+\hat P f}}-\hat P\bar{f}-\hat P \left [\bar{f}{f_u}\right ]=0
\end{equation}
with $f$ and $\bar f$ given by equations \e{fdef} and \e{fdefbar}, respectively.

\subsection{Projection in $\zeta$ plane}
\label{sec:Projectionzeta}

The projector $\hat P$ \e{Projectordef} is defined in terms of the
independent variable $u$. Using  equation \e{eqn:stokes22}
together with the definition \e{fdef} suggests to switch from $u$
into the independent variable $\zeta $.  To identify how to compute
$\hat P$  in complex  $\zeta$-plane,  we start from the Fourier
series \e{ffourier}, \e{fpm0}  in variable $u$  and make a change of
variable \e{zetadef} (assuming that $-\pi\le u\le\pi$ and
$\zeta\in{\mathbb{R}})$ as follows
\begin{align} \label{fouruzeta}
f(u)=f(\zeta)=\sum\limits_{k=-\infty}^{\infty}f_ne^{\I ku}=\sum\limits_{k=-\infty}^{\infty}f_n\exp\left[
2\I k\arctan{\zeta} \right ]=\sum\limits_{n=-\infty}^{\infty}f_k
\left( \frac{\zeta-\I }{\zeta+\I } \right )^k (-1)^k,
\end{align}
where we abuse notation by assuming that $\tilde f(\zeta)\equiv f(u) $
and removing $\tilde ~$ sign. Equations  \e{Projectordef}, \e{fprojected} and
\e{fouruzeta} imply that  $\hat P$ removes all Fourier harmonics
with positive $n$ and replaces the zeroth harmonic $f_0$ by $f_0/2$ as
follows
\begin{align} \label{projectoru}
\hat P f(u)=\sum\limits_{n=-\infty}^{\infty}f_k\hat
Pe^{\I ku}=\frac{f_0}{2}+\sum\limits_{k=-\infty}^{-1}f_k\exp\left[
\I k2\arctan{\zeta} \right ]\nonumber
\\=\frac{f_0}{2}+\sum\limits_{n=-\infty}^{-1}f_k \left(
\frac{\zeta-\I }{\zeta+\I } \right )^k (-1)^k.
\end{align}

Consider a particular case $f(u)=\frac{1}{\zeta-\I \chi}$,
$\chi\in\mathbb{R}$ and $\chi\ne 0.$ We calculate $f_k$  by
equation \e{ffourier} and   \e{fouruzeta}  through the change of
variable \e{zetadef} implying $du=\frac{2}{\zeta^2+1}d\zeta$ as
follows
\begin{align} \label{uzetaint}
f_{-k}=\frac{1}{2\pi}\int^\pi_{-\pi}f(u)e^{\I ku}du=\frac{1}{2\pi}\int^\infty_{-\infty}\frac{1}{\zeta-\I \chi}
\left( \frac{\zeta-\I }{\zeta+ \I } \right )^k
(-1)^k\frac{2}{\zeta^2+1}d\zeta.
\end{align}
Assuming $k\ge 0$ and closing the complex integration contour in the upper
half-plane of $\zeta$ we obtain that
\begin{align} \label{aminun}
f_{-k}=\I  \left( \frac{\chi- 1}{\chi+ 1} \right )^k
(-1)^k\frac{2}{-\chi^2+1}\theta(\chi)+\delta_{k,0}\frac{1}{\I -\I \chi}.
\end{align}
For the zeroth harmonic $f_0$,  equation  \eqref{aminun} results in
\begin{align} \label{aminun0}
f_{0}=\frac{\I\,\text{sign}(\chi)}{1+\chi\,\text{sign}(\chi)},
\end{align}
where $\text{sign}(\chi)=1$ for $\chi>0$ and $\text{ \
\,sign}(\chi)=-1$ for $\chi<0.$

Using now equations \eqref{projectoru}, \eqref{aminun} and \eqref{aminun0} we
find that
\begin{align} \label{aminuncont}
\hat P
\frac{1}{\zeta-\I \chi}=\frac{-\I\,\text{sign}(\chi)}{2[1+\chi\,\text{sign}(\chi)]}+\sum\limits_{k=0}^{\infty}\left
[ \I  \left( \frac{\chi- 1}{\chi+ 1} \right )^k
(-1)^k\frac{2}{-\chi^2+1}\theta(\chi)+\delta_{k,0}\frac{1}{\I -\I \chi}\right
] \nonumber \\ \times\left( \frac{\zeta+\I }{\zeta- \I } \right )^k
(-1)^k=\frac{1}{\zeta-\I \chi}\theta(\chi)+\frac{1}{\I -\I \chi}\theta(-\chi)-\frac{\I\,\text{sign}(\chi)}{2[1+\chi\,\text{sign}(\chi)]},
\end{align}
where  $\theta(\chi)=1$ for $\chi>0$ and $\theta(\chi)=0$ for
$\chi<0$.

In a similar way, for $f(\zeta)=\dfrac{1}{(\zeta - \I \chi)^2}$ we
find from the series \e{fouruzeta}  that
\begin{align} \label{uzetaint2}
f_{-k}=\frac{1}{2\pi}\int^\infty_{-\infty}\frac{1}{(\zeta-\I \chi)^2}
\left( \frac{\zeta- \I }{\zeta+\I } \right )^k
(-1)^k\frac{2}{\zeta^2+1}d\zeta, \quad k\ge 0.
\end{align}
Closing the complex integration contour in the upper half-plane of
$\zeta$ one obtains from  equation \e{uzetaint2} that
\begin{align} \label{aminun2}
f_{-k}=\left .\I  \frac{d}{d\zeta}\left( \frac{\zeta-\I }{\zeta+\I }
\right )^k (-1)^k\frac{2}{\zeta^2+1}\theta(\chi)\right
|_{\zeta=i\chi}+\delta_{k,0}\frac{1}{(\I -\I \chi)^2}, \quad k\ge 0
\end{align}
and
\begin{align} \label{aminun20}
f_0=-\frac{1}{[1+\chi\,\text{sign}(\chi)]^2}.
\end{align}

Taking a sum over $k$ in  equation  \eqref{projectoru}, using equations
\e{aminun2} and  \e{aminun20}  and we find that
\begin{align} \label{aminuncont2}
\hat P \dfrac{1}{(\zeta -\I \chi)^2}=\frac{1}{(\zeta-\I \chi)^2}\theta(\chi)+\frac{1}{(\I -\I \chi)^2}\theta(-\chi)+\frac{1}{2[1+\chi\,\text{sign}(\chi)]^2}.
\end{align}

\subsection{Integral representation of the equation for Stokes wave}
\label{sec:Integralrepreseantation}

Using  equations  \e{fdef}, \e{aminuncont} and  \e{aminuncont2} we
obtain the following projections in terms of $\rho(\chi)$:
\begin{align} \label{Pbarf}
\hat P{f}= \int\limits_{\chi_c}^{1} \dfrac{\rho(\chi) d\chi}{\zeta -
i\chi}-\int\limits_{\chi_c}^{1} \frac{\I\rho(\chi)
d\chi}{2(1+\chi)}, \qquad \hat P\bar{f}=- \int\limits_{\chi_c}^{1}
\dfrac{\I\rho(\chi) d\chi}{2(1+\chi)}.
\end{align}

We now find $\hat P \left [\bar{f}{f_u}\right ]$ used in
\eqref{eqn:stokes22}. Equation  \e{zetadef}  results in the
following expression
\begin{align} \label{fbarfu}
\bar f f_u=\frac{\zeta^2+1}{2}\bar f f_{\zeta} =-\frac{\zeta^2+1}{2}
\int\limits^1_{\chi_c}\int\limits^1_{\chi_c}
\dfrac{\rho(\chi')\rho(\chi'')d\chi'
d\chi''}{(\zeta+\I \chi')(\zeta-\I \chi'')^2}.
\end{align}
We perform the partial fraction decomposition of the integrand of
\eqref{fbarfu} as follows
\begin{align}\label{apart1}
-\dfrac{\zeta^2+1}{2(\zeta + \I \chi')(\zeta - \I \chi'')^2} =
\dfrac{1}{\zeta+\I \chi'}  \dfrac{1-\chi'^2}{2(\chi' + \chi'')^2}+
\dfrac{1}{\zeta -\I \chi''}  \dfrac{-1-2\chi'\chi''-\chi''^2}{2(\chi'
+ \chi'')^2}\nonumber \\+ \dfrac{1}{(\zeta - \I \chi'')^2}
\dfrac{\I (1-\chi''^2)}{2(\chi' + \chi'')}.
\end{align}
and apply the projector $\hat P$ to \eqref{apart1} which gives with
the use of \eqref{aminuncont} and \eqref{aminuncont2} the following
expression:
\begin{align}\label{Pffu}
\hat P \left [\bar{f}{f_u}\right
]=\int\limits^1_{\chi_c}\int\limits^1_{\chi_c} \left
[\dfrac{1}{\I +\I \chi'}  \dfrac{1-\chi'^2}{4(\chi' + \chi'')^2}+\left (
\dfrac{1}{\zeta - \I \chi''}-\dfrac{\I}{2(1 +\chi'')} \right ) \dfrac{-1-2\chi'\chi''-\chi''^2}{2(\chi' + \chi'')^2} \right .\nonumber\\
\left.+\left ( \dfrac{1}{(\zeta - \I \chi'')^2}+\dfrac{1}{2(1
+\chi'')^2} \right ) \dfrac{\I (1-\chi''^2)}{2(\chi' + \chi'')}
\right ]\rho(\chi')\rho(\chi'')d\chi' d\chi''.
\end{align}

The other nonlinear term in equation \eqref{eqn:stokes22} has the following
integral form
\begin{align}\label{ffu}
ff_u =\frac{1}{2}(1+\zeta^2)f f_\zeta= -\frac{1}{2}(1+\zeta^2)
\int\limits^1_{\chi_c} \dfrac{\rho(\chi')d\chi'}{(\zeta -\I \chi')}
\int\limits^1_{\chi_c} \dfrac{\rho(\chi'') d\chi''}{(\zeta -
\I \chi'')^2}.
\end{align}

The constant $y_b$ is determined from  equation \e{yxucondition} as
follows
\begin{equation} \label{yxucondition2}
\int\limits^{\pi}_{-\pi} y(1+\tilde x_u)
du=\int\limits^{\infty}_{-\infty}\left [y_b+\frac{(f-\bar
f)}{2\I}\right ]\left [1+\frac{(1+\zeta^2)}{4}(f_\zeta+\bar
f_\zeta)\right ]\frac{2d\zeta}{1+\zeta^2}=0,
\end{equation}
which results using equation \e{fdef}  in the following equation
\begin{align} \label{massconservationinf}
y_b=-
\int\limits^1_{\chi_c}\int\limits^1_{\chi_c}\frac{\rho(\chi')\rho(\chi'')d\chi'd\chi''}{2(\chi'+\chi'')^2}-\int\limits^1_{\chi_c}\frac{\rho(\chi')d\chi'}{1+\chi'}.
\end{align}
Equation \e{massconservationinf}  allows to find $y_b$ from a given
$\rho(\chi).  $ This equation also provides a convenient tool to
estimate the accuracy of recovering $\rho(\chi)$ by Pad\'e
approximation. For that one compares  the numerical value of  $y_b$
obtained from the Stokes solution in Part I
\citep{DyachenkoLushnikovKorotkevichPartIStudApplMath2016} with the
result of the direct numerical calculation of right-hand side
(r.h.s.) of  equation \e{massconservationinf} with $\rho(\chi)$
obtained from Pad\'e approximation in Part I (all these numerical
values are given in tables of Part I
\citep{DyachenkoLushnikovKorotkevichPartIStudApplMath2016}, through
the electronic attachment to Ref.
\citet{DyachenkoLushnikovKorotkevichPartIArxiv2015} and at the web
link~\citet{PadePolesList}).

Integrating equation \e{fdef} in $u$ over  $2\pi$-period one obtains the zero Fourier harmonic $y_0$  of $y(u)$ as follows%
\begin{equation} \label{y0throughyb}
y_0=y_b+\int\limits^1_{\chi_c}\frac{\rho(\chi')d\chi'}{1+\chi'}.
\end{equation}

Requiring that equations \e{fdef}-\e{eqn:stokes22}, \e{Pbarf},
\e{Pffu}, \e{ffu} and \e{massconservationinf} are satisfied for
$-\infty<\zeta<\infty$ we obtain  a  system of equations to
find the density $\rho(\chi)$ along the branch cut for each $c$.
That system has a form of nonlinear integral equation for  the
unknown function $\rho(\chi).$ Taking the limit $\zeta\to\infty $ in  that system
 results
 in the following compact expression
\begin{align} \label{invzeta0}
&\frac{c^2}{2}\int\limits^1_{\chi_c}\rho(\chi')d\chi'+2\left [-
\int\limits^1_{\chi_c}\int\limits^1_{\chi_c}\frac{\rho(\chi')\rho(\chi'')d\chi'd\chi''}{2(\chi'+\chi'')^2}-\int\limits^1_{\chi_c}\frac{\rho(\chi')d\chi'}{1+\chi'} \right ]\nonumber \\
&\times \left
[1-\frac{1}{2}\int\limits^1_{\chi_c}\rho(\chi''')d\chi''' \right
]+\int\limits^1_{\chi_c}\frac{\rho(\chi')d\chi'}{1+\chi'}+
\int\limits^1_{\chi_c}\int\limits^1_{\chi_c}\frac{(1-\chi')\rho(\chi')\rho(\chi'')d\chi'd\chi''}{2(\chi'+\chi'')^2}=0,
\end{align}
which can be used to find $c$ from the given $\rho(\chi).$

\subsection{Numerical solution for Stokes wave based on the integral representation}
\label{sec:Integralrepreseantationnumerical}

To solve the system  \e{fdef}-\e{eqn:stokes22}, \e{Pbarf}, \e{Pffu},
\e{ffu} and \e{massconservationinf} numerically we use the approximation of the integral in equation  \eqref{branchcutdensity} by the following numerical quadrature
\begin{equation}
\label{approx_cut0}  f(u)= \tilde z(\zeta) - \I y_b = \int\limits_{\chi_c}^{1} \dfrac{\rho(\chi')d\chi'}{\zeta - \I \chi'} \simeq
\sum\limits_{j = 1}^{N} \dfrac{\gamma_j}{\zeta - \I \chi_j},
\end{equation}
which has a form of Pad\'e approximation at the
discrete set of points $\chi_c<\chi_1<\chi_2<\ldots<\chi_N<1\   $  with weights $\gamma_j , \ j=1,2,\ldots
N.$ Then the analysis of Sections \ref{sec:Projectionzeta} and \ref{sec:Integralrepreseantation}
 with equation   \eqref{branchcutdensity} replaced by  the approximation
 \e{approx_cut0} is carried out in exactly the same way as in equations \e{fdef}-\e{massconservationinf} with each time $\rho(\chi)d\chi$ and $\chi$ replaced by $\gamma_j$ and $\chi_j$, respectively. Also integrals are replaced by the summations. It results in the  discrete versions of these equations including
\begin{align} \label{Pbarfdiscrete}
\hat P{f}= \sum\limits^{N}_{j=1} \dfrac{\gamma_j}{\zeta -
i\chi}-\sum\limits^{N}_{j=1} \frac{\I\gamma_j}{2(1+\chi)}, \qquad \hat P\bar{f}=- \sum\limits^{N}_{j=1}
\dfrac{\I\gamma_j}{2(1+\chi)},
\end{align}
\begin{align}\label{ffudiscrete}
ff_u = -\frac{1}{2}(1+\zeta^2)
 \sum\limits^{N}_{j'=1} \dfrac{\gamma_{j'}}{(\zeta -\I \chi_{j'})}
 \sum\limits^{N}_{j''=1} \dfrac{\gamma_{j''}}{(\zeta -
\I \chi_{j''})^2}
\end{align}
and
\begin{align}\label{Pffudiscrete}
\hat P \left [\bar{f}{f_u}\right
]=\sum\limits^{N}_{j'=1}\sum\limits^{N}_{j''=1} \left
[\dfrac{1}{\I +\I \chi_{j'}}  \dfrac{1-\chi_{j'}^2}{4(\chi_{j'} + \chi_{j''})^2}+\left (
\dfrac{1}{\zeta - \I \chi_{j''}}-\dfrac{\I}{2(1 +\chi_{j''})} \right )
 \right .\nonumber\\
\left.\times \dfrac{-1-2\chi_{j'}\chi_{j''}-\chi_{j''}^2}{2(\chi_{j'} + \chi_{j''})^2}+\left ( \dfrac{1}{(\zeta - \I \chi_{j''})^2}+\dfrac{1}{2(1
+\chi_{j''})^2} \right ) \dfrac{\I (1-\chi_{j''}^2)}{2(\chi_{j'} + \chi_{j''})}
\right ]\gamma_{j'}\gamma_{j''}.
\end{align}
Also equation \e{massconservationinf} is replaced in the same discrete approximation by the following equation
\begin{align} \label{massconservationinfPade}
y_b=-\sum\limits^{N}_{j'=1}\sum\limits^{N}_{j''=1}\frac{\gamma_{j'}\gamma_{j''}}{2(\chi_{j'}+\chi_{j''})^2}-\sum\limits^{N}_{j'=1}\frac{\gamma_{j'}}{1+\chi_{j'}}.
\end{align}

Choosing numerical values $\gamma_j, \ \chi_j, \ j=1,2,\ldots ,N$
from Pad\'e approximants of Part I
\citep{DyachenkoLushnikovKorotkevichPartIStudApplMath2016} (these
approximants  are also available  through the electronic attachment
to Ref. \citet{DyachenkoLushnikovKorotkevichPartIArxiv2015} and at
the web link~\citet{PadePolesList}) we checked that equation
\e{eqn:stokes22} (together with equations
\e{Pbarfdiscrete}-\e{massconservationinfPade}) is valid for each
value of $H/\lambda$ with the same numerical precision  as the
precision (at least $10^{-26}$) of the Stokes solutions of Part I
\citep{DyachenkoLushnikovKorotkevichPartIStudApplMath2016}. Values
of $N$ in Part I range between tenths for moderates values of
$H/\lambda$ up to $N=92$ for the highest Stokes wave considered
(given by Table 4 in Part I
\citep{DyachenkoLushnikovKorotkevichPartIStudApplMath2016}). These
moderate numbers is in sharp contrast with the large number $M$ of
Fourier modes required for constructing these solutions with the
same precision ($M\simeq1.3\cdot 10^8$ for the highest Stokes wave
considered in Table 4 of Part I). An explanation for that dramatic
difference between required numerical values of $M$ and $N$ follows
from Part I. It was found in Part I that the error of Fourier method
scales as $\propto\exp{(-2\chi_cM)}$ while the error for Pad\'e
approximation of  Part I is $\propto \exp{(-c_1\chi_c^{1/6}M)}, \
c_1\sim1$.  It suggests that solving equations \e{eqn:stokes22},
\e{Pbarfdiscrete}-\e{massconservationinfPade} for numerical values
of  $\gamma_j, \ \chi_j, \ j=1,2,\ldots ,N$ is the attractive
alternative to the numerical methods of Part I.

To solve equations  \e{eqn:stokes22}, \e{Pbarfdiscrete}-\e{massconservationinfPade} numerically, we aim to approximately satisfy
equation   \e{eqn:stokes22} at the discrete set of points     $-\infty<\zeta=\zeta_i<\infty,\  i=1,2,\ldots ,M_1. $  It results
in the nonlinear algebraic system of equations to find  $\gamma_j, \ \chi_j, \ j=1,2,\ldots, N$. That system is overdetermined
(see e.g. Ref. \citet{WilkeningYuComputSciDisc2012} as the example of using of overdetemined systems for simulating water waves)
provided we choose  $M_1>2N$ but it can be solved in least square sense (by minimizing the sum of squares of the left-hand side
(l.h.s.) of  equation \e{eqn:stokes22} taken over points $\zeta=\zeta_i, \ i=1,2,\ldots, M_1).$ However, the difficulty in such
most straightforward approach is in extreme ill-conditioning of the resulting algebraic system mainly because of denominators
containing large powers of $\chi_j$ clearly seen if we try to bring equation   \e{eqn:stokes22} to the common denominator. We
bypass that difficulty  here by providing the explicit procedure to find the appropriate values of  $\chi_j, \ j=1,2,\ldots N$
for each $\chi_c$ (see the description of that procedure below in this Section) and only after that we solve equations
\e{eqn:stokes22}, \e{Pbarfdiscrete}-\e{massconservationinfPade} for unknowns $\gamma_j, \ j=1,2,\ldots ,N$ at the discrete set of
points $-\infty<\zeta=\zeta_i<\infty,\  i=1,2,\ldots ,M_1. $ Then the resulting system is the cubic polynomial  in  $\gamma_j, \
j=1,2,\ldots N$. That system is still moderately ill-conditioned but that difficulty is easily overcome by choosing $M_1$ large
enough with Newton's iterations used to find numerical values of $\gamma_j, \ j=1,2,\ldots N,$ thus forming least-square-Newton
(LSN) algorithm. E.g., for $H/\lambda=0.1387112446\ldots$ (corresponds to $\chi_c=3.0056373876\ldots\cdot10^{-3}$, see also table
\ref{tab:H.chic} of Appendix \ref{sec:TablesStokesWaveschic} for details on numerical Stokes waves) we found that it is
sufficient to use $M_1=800$ and $N=51$ to achieve $10^{-19}$ accuracy for Stokes wave. For steeper Stokes waves with
$H/\lambda=0.1401109676\ldots$ ($\chi_c=6.99513864872\ldots\cdot 10^{-4}$) and $H/\lambda=0.1408682599\ldots$
($\chi_c=5.6590609636\ldots\cdot 10^{-5}$)  we found that using $M_1=1600, \ N=61$ and  $M_1=10^4, \ N=78$  allow to achieve
$10^{-18}$ and  $10^{-19}$  accuracy, respectively. Here values of $N$ were chosen the same as for the respective Stokes wave in
Part I while $M_1$ is by a factor $\sim 80$ smaller than $M=65536$  in the first case and by a factor  $\sim 200$  smaller than
$M=2097152$ in the third case (values of $M$ are given  in Part I,  through the electronic attachment to Ref.
\citet{DyachenkoLushnikovKorotkevichPartIArxiv2015} and at the web link~\citet{PadePolesList}). In these examples, using the
symmetry of Stokes wave, the points $\zeta_j$ were chosen to have nonnegative values with the first  300 points uniformly spaced
as $\zeta_i=(i-1)2\pi/M$, $ \ i=1, \ldots, 300$ and the remaining $M_1-300$ points uniformly (in $u$) spanning the remaining
interval of positive values of $\zeta.$   After values of $\gamma_j, \ j=1,2,\ldots N$ are found from LSN algorithm, equation
\e{approx_cut0} provides Pad\'e approximation for Stokes wave at the entire real line of $\zeta.$  Then one can use the results
of Section \ref{sec:chicmatching} to find the high precision numerical approximation of $\chi_c,$ which completes the current
step in $H/\lambda$ (or equivalently the current step in $\chi_c$). These step are repeated to gradually increase  $H/\lambda$
(or equivalently decrease $\chi_c$) by changing the velocity parameter $c$ to span the desired range of Stokes waves.

The procedure to find the grid    $\chi_j, \ j=1,2,\ldots N$ at each
step is the following. Assume that
$\chi_c<\chi_1<\chi_2<\ldots\chi_{N-1}<\chi_N<1$ and  $\chi_c \ll
1$. We use  the property of Stokes wave that $\rho(\chi)$ changes a
little vs. a change of $H/\lambda$ for $\chi\gg \chi_c$ provided
$\chi_c \ll 1$.  It implies that      $\chi_j$ can be chosen
independently on $\chi_c$ for all $j$ such that       $\chi_j\gg
\chi_c$. In numerical examples above we chose numerical values in
the range        $\chi_j\gg \chi_c$ from Pad\'e data for Stokes wave
with $H/\lambda=0.1409700957\ldots$  obtained in Part I
\citep{DyachenkoLushnikovKorotkevichPartIStudApplMath2016}. Also  the grid  $\chi_j$   can be chosen from the grid obtained at the previous
step (with a previous smaller value of $\chi_c$).

We  now consider the construction of grid for smaller values of $\chi_j.$ If we assume a power law singularity $\rho(\chi)\propto(\chi-\chi_c)^\alpha, \alpha>0$  and consider the limit $\zeta\to \I\chi_c$ in equation   \eqref{branchcutdensity}, then the transformation to a new integration variable $t=(\chi-\chi_c)^\alpha$ removes the singularity from the integrand in equation   \eqref{branchcutdensity}. The uniform grid $t_j=j\Delta t, \ j=1,2,\ldots, \ \Delta t=const $ in $t$ is the natural choice to use for the integration in the variable  $t$. The corresponding grid in $\chi$ is given by
\begin{equation} \label{alphagrid}
\chi_j-\chi_c=t_j^{1/\alpha}=j^{1/\alpha}\Delta t^{1/\alpha}.
\end{equation}

Stokes wave has the square root singularity at $\zeta=\I\chi_c  $
with  the expansion
\begin{align} \label{zetachicserp0}
\tilde z(\zeta)-\I y_b=f(\zeta)=\sum\limits_{j=0}^\infty
\I e^{\I j\pi/4}a_{j}(\zeta-\I\chi_c)^{j/2}, \quad
\end{align}
where $a_j$ are real constants (see Part I \citep{DyachenkoLushnikovKorotkevichPartIStudApplMath2016} as well as Sections \ref{sec:SeriesexpansionsStructureRiemannsurface} and \ref{sec:squarerootsingularities} below for the justification of that expansion). It implies (see Part I) the square root singularity for the density  $\rho(\chi)\propto (\chi-\chi_c)^{1/2}$  in the integrand of  equation   \eqref{branchcutdensity}. Using equation \e{alphagrid} with $\alpha=1/2$ one then obtains that%
\begin{equation} \label{chi1p2}
\chi_j-\chi_c=\Delta t^2j^2, \ j\sim 1,
\end{equation}
which is in the excellent agreement with numerical values of  $\chi_j$ obtained in Part I provided $\Delta t^2\sim 0.01\chi_c$.

In the range $\chi_c\ll \chi \ll 1$, the density $\rho(\chi)\propto
\chi^{2/3}$   is well approximated by the density $\rho(\chi)\propto
\chi^{2/3}$    of  the limiting Stokes wave \e{w2p3} as shown in
Figure 8 of Part I. Using equation \e{alphagrid} with $\alpha=2/3,$
one  obtains that

\begin{equation} \label{chi3p2}
\chi_j=cj^{3/2},
\end{equation}
where $c$ is the positive constant and $j\gg 1 $ such that  $\chi_j\ll 1$. We additionally have to approximate the transition between two scalings \e{chi1p2} and \e{chi3p2} at the intermediate values of $j$. Exploring fits of $\chi_j$ vs. $j$ for multiple sets of numerical data of Part I we found that  a satisfactory fit (including the required transition) is given by the linear combination of the scaling  \e{chi1p2} and \e{chi3p2} superimposed with the exponential growth in $j$ as follows%
\begin{equation} \label{chijcombined}
\chi_j=\chi_c \left [c_1j^2+c_2j^{3/2}e^{c_3 j} \right ],
\end{equation}
where the positive fitting constant $c_1, \, c_2$ and $c_3$ changes slowly with $\chi_c$ (change of $\chi_c$ in 5 orders of magnitude results in change of these constants by less than 50\%).

Based on these observations we implemented the following procedure to find numerical values of  $c_1, \, c_2$ and $c_3$ for each value of $\chi_c.$ We  choose $N$ from the previous step (with the previous value of $\chi_c$). (If performing the current step we are not able to  reach the desired precision with the increase of $M_1$, i.e. LSN algorithm would not converge to the prescribed tolerance, e.g. $10^{-16}$, then $N$ has to be increased by 1 and the current step restarted from the beginning). Next we choose $j_{match}$ from the values $\chi_j$ of   Table 4 of Part I (or from the grid obtained at the previous step in $\chi_c$) such that $\chi_{j_{match}}/\chi_c\sim 100$ which well ensures the required condition $\chi_j\gg \chi_c.$ (For larger values of $\chi_c$ one can use a smaller value of  $\chi_{j_{match}}/\chi_c$ to make sure that $\chi_{j_{match}}\ll  1$. E.g. for the case  $H/\lambda=0.1387112446\ldots$  (first numerical case mentioned above in this section) we choose   $\chi_{j_{match}}/\chi_c\sim 34.)$     After choosing the value of $j_{match}$, we perform 4th order interpolation of $\chi_j$ as the function of $j$ and find values of the first and second derivatives, $\chi_j'$ and $\chi_j''$, of that interpolant at $j=j_{match}$. We use these 3 numerical values $\chi_{j_{match}}$,  $\chi_{j_{match}}'$ and $\chi_{j_{match}}''$ to find the numerical values of $c_1, \ c_2$ and $c_3$ by matching the corresponding values of equation \e{chijcombined} and its two derivatives at $j=j_{match}.$ Then equation \e{chijcombined} provides the numerical values of $\chi_j, \ j=1,2, \ldots,j_{match}$ completing
the construction of the numerical grid  $\chi_j, \ j=1,2, \ldots,N$ for the current step in $\chi_c.$ (If any of the constants $c_1,c_2$ or $c_3$ turns negative then one has to decrease $\chi_{j_{match}}/\chi_c$ to avoid that but in our numerical examples we experienced such problems only if $\chi_{j_{match}}/\chi_c$ was chosen $\gtrsim10^3).$ Then LSN algorithm is used as described above.

The efficiency of the grid    $\chi_j, \ j=1,2,\ldots
N$ thus obtained requires a good initial estimate of $\chi_c$ with the relative accuracy $\sim 10^{-3}$. It is achieved by a gradual increase of $H/\lambda$ (decrease of $\chi_c$) at multiple previous steps of LSN algorithm. Values of  $\chi_c$ are found at each previous step with high precision by the procedure of Section \ref{sec:chicmatching}. The polynomial extrapolation of $\chi_c$ to the current step  is performed to reach the needed relative accuracy $\sim 10^{-3}$. Note that the numerical detection of the incorrect value of $\chi_c$ prediction is straightforward because    it would cause the oscillations of  $\gamma_j$  (with changing of its sign) around several smallest values of the index $j=1,2,3,\ldots$.

We would like to stress the difference of LSN algorithm of this Section compare with the method of Part I
\citep{DyachenkoLushnikovKorotkevichPartIStudApplMath2016}. Stokes wave was obtained in Part I by using the Fourier series
representation of the solution combined with Newton-Conjugate-Gradient iterations method. After that the resulting solution was
approximated at the real line of $\zeta$ by a numerically stable version of the Pad\'e algorithm. Thus Pad\'e approximants of
Part I were only the auxiliary  tool to compactly represent the result of calculation of Stokes waves. In contrast, in this
Section we completely bypass the Fourier series representation and numerically solve integral equations \e{eqn:stokes22},
\e{Pbarfdiscrete}-\e{massconservationinfPade} directly by Pad\'e approximants. The cost of the approach  of this Section is that
instead of $2N$ free parameters $\chi_j, \gamma_j, \ N=1,2,\ldots,N $ of Pad\'e approximats of Part I, we now have only $N$ free
parameters  $ \gamma_j, \ N=1,2,\ldots,N, $    while the values of$\chi_j,  \ N=1,2,\ldots,N $ are fixed by the  grid algorithm
described previously in this section. It means that to achieve the same precision we need to approximately double the value of
$N$ compare with Part I. This is however very moderate cost compare with  the Fourier  method of Part I.

In conclusion, in this Section we demonstrated the performance LSN algorithm for several values of $\chi_c$ which were previously explored in Part I by the Fourier method.  We expect that much smaller values of $N,$ required for LSN algorithm compare with Fourier method, will allow to find Stokes waves for much smaller values of $\chi_c$ than achievable by Fourier method of Part I. In addition, equation \e{invzeta0} can be used to exclude $c$ from the system  allowing to gradually increase $H/\lambda$ (or equivalently decrease $\chi_c$) thus avoiding the problem of nonmonotonic dependence of $c$ on $H/\lambda$ encountered in Part I.  The  detailed practical realization of that limit of smaller $\chi_c$ is
beyond the scope of this paper.


\section{Alternative form for the equation of  Stokes wave}

\label{sec:AlternativeStokeswave}

The equation
for Stokes wave can be written in a form which is alternative to equation    \e{stokes_wave}   as follows%
\begin{equation} \label{Stokeszuzubar}
y=-\frac{\I}{2}(\tilde z-\bar {\tilde {z}})=-\frac{\I}{2}( z-\bar { {z}})=\frac{c^2}{2}\left (1-\frac{1}{|z_u|^2}\right ).
\end{equation}
Appendix \ref{sec:Stokeswaveequivalnce}  shows the equivalence of both forms of equations  \e{stokes_wave} and \e{Stokeszuzubar} for Stokes wave. Also  Appendix \ref{sec:restmovingframe}   discusses differences in derivation of equations   \e{stokes_wave} and \e{Stokeszuzubar}  from
basic equations of the potential flow of ideal fluid with free surface.
Different versions of equation   \e{Stokeszuzubar}
 (up to trivial scaling of parameters and shift of $y$ by different constants)
were used  by
\citet{MalcolmGrantJFM1973LimitingStokes}, \citet{Williams1981},
\citet{Plotnikov1982} and \citet{TanveerProcRoySoc1991}.

Transforming equation \e{Stokeszuzubar} into the variable  $\zeta$   \e{zetadef}   results in
\begin{equation} \label{Stokeszuzubarzeta}
\tilde z-\bar {\tilde {z}}={\I c^2}\left (1-\frac{4}{(1+\zeta^2)^2|z_\zeta|^2}\right ).
\end{equation}

Solving equation \e{Stokeszuzubar} for $z_u$, one obtains that
\begin{equation} \label{Stokeszuzubarzetaexplicit}
z_u=\frac{c^2}{\bar z_u[\I(z-\bar z)+c^2]}
\end{equation}
which is the nonlinear ODE provided $\bar z_u$ is known. In a similar way, equation \e{Stokeszuzubarzeta} results in
\begin{equation} \label{Stokeszuzubarexplicit}
z_\zeta=\frac{4}{(1+\zeta^2)^2}\frac{c^2}{\bar z_\zeta[\I(z-\bar z)+c^2]}.
\end{equation}

Equations \e{Stokeszuzubarzetaexplicit} and
\e{Stokeszuzubarexplicit} can be considered as ODEs for $z(u)$ and
$z(\zeta)$, respectively, if $\bar z$ is the known function.
Then solving ODE provides a convenient tool to study the analytical
properties of Stokes wave in different sheets of Riemann surface of
$z$.

\section{Asymptotic of Stokes wave at $Im(w)\to +\infty$ and jump at branch cut}
\label{sec:AsymptoticStokeswave}

An asymptotical solution of Stokes wave in the limit $Im(w)\to +\infty$ is obtained from equation  \e{Stokeszuzubarzetaexplicit}
as follows. Equation \e{fminus} implies  the exponential convergence $\propto e^{-\I w}$ of $\tilde z(w)$    to its zeroth
Fourier harmonic, $\tilde z(w)\to \I y_0$     for $Im(w)\to -\infty$.  Here $y_0$ is determined by the   mean-zero elevation
condition \e{yxucondition} and is given by equations \e{massconservationinf} and \e{y0throughyb}. Respectively, $\bar{ \tilde
{z}}(w)$ converges exponentially to $-\I y_0$ for $Im(w)\to \infty$. Then $\bar z_u$ and $\bar z$ in equation
\e{Stokeszuzubarzetaexplicit} can be replaced by $1$ and $-\I y_0$, respectively in that limit resulting in
\begin{equation} \label{Stokeszuzubarzetaexplicitasymptotics}
1+\tilde z_u=\frac{c^2}{\I(\tilde z+\I y_0)+c^2}, \qquad    Im(w)\to \infty.
\end{equation}

Integrating equation \e{Stokeszuzubarzetaexplicitasymptotics} in the upper right quadrant  $w\in\mathbb{C}^+, \ Re(w)>0$  for $v\gg 1$, one obtains that %
\begin{equation} \label{zplus}
\tilde z(w)-\I c^2\ln{\left [ \tilde z(w)+\I y_0 \right ]}=-w+c_+,
\end{equation}
where $c_+$ is the constant. A similar integration in the upper left quadrant  $w\in\mathbb{C}^+, \ Re(w)<0$, for $v\gg 1$ results in %
\begin{equation} \label{zminus}
\tilde z(w)-\I c^2\ln{\left [ \tilde z(w)+\I y_0 \right ]}=-w+c_{-},
\end{equation}
where $c_-$ is the  constant.

Taking $w=\pi+\I v$   in equation \e{zplus} and  $w=-\pi+\I v$    in equation \e{zminus} together with the periodicity condition $\tilde z(\pi+\I v) =\tilde z(-\pi+\I v)$ result in the condition for constants $c_+$ and $c_-$ as follows
\begin{equation} \label{cpm}
c_--c_+=-2\upi.
\end{equation}

Exponents of equations \e{zplus} and \e{zminus} are similar to the Lambert $W$-function. Solving these equations in the limit $v\to \infty$ (see e.g. Refs. \citet{LushnikovDyachenkoVladimirovaNLSloglogPRA2013,DyachenkoLushnikovVladimirovaKellerSegelNonlinearity2013} for details on a similar technique)  one obtains that
\begin{align} \label{zplusminusasymp2}
\tilde z(w)=-w+c_{\pm}+\I c^2\ln{\left [  -w+c_{\pm} +\I y_0 \right ]}-\frac{c^4\ln{\left [  -w+c_{\pm} +\I y_0 \right ]}}{ -w+c_{\pm} +\I y_0}\nonumber \\
+O\left( \frac{\ln{\left [  -w+c_{\pm} +\I y_0 \right ]}}{ -w+c_{\pm} +\I y_0}
 \right )^2,
\end{align}
where a use of  $c_+$ and $c_-$ assumes that $Re(w)>0$ and $Re(w)<0$, respectively. If equation  \e{Stokeszuzubarzetaexplicit} is used instead of the reduced equation  \e{Stokeszuzubarzetaexplicitasymptotics} in derivation of equation \e{zplusminusasymp2}, then an additional exponentially small error term $O(e^{-v}/v$) appears in r.h.s. of equation \e{zplusminusasymp2}.
The two leading order terms $-w+c_{\pm}$ and $\I c^2\ln{\left [  -w+c_{\pm} +\I y_0 \right ]}$ in r.h.s. of equation \e{zplusminusasymp2} are similar to equation  (2.22) of Ref. \citet{TanveerProcRoySoc1991}, where these terms were derived in somewhat similar procedure to the derivation of equation \e{zplusminusasymp2}.

One concludes from equation  \e{zplusminusasymp2} that $z(w)$ has a complex singularity at  $z=\infty$ which involves  logarithms with the infinite number of sheets of Riemann surface. Full analysis of that singularity requires to study next order terms in   equation \e{zplusminusasymp2} which is beyond the scope of this paper.

Taking an additional limit $w=\I v \pm\epsilon$, $\epsilon>0, \ \epsilon\to 0  $  in equation \e{Stokeszuzubarzetaexplicitasymptotics}, using the condition \e{cpm}  and expanding in $v\gg 1,$ one obtains the jump at the branch cut
\begin{equation} \label{zjump}
z(\I v -0)-z(\I v +0)=-2\upi+\frac{2\upi  c^2}{  v}+O(v^{-2}),
\end{equation}
where the branch cut $v\in [\I v_c,\I\infty]$ is crossed in counterclockwise direction.

According to equation \e{jumprho1}, the jump  \e{zjump} is related to the density $\rho(\chi)$  \e{branchcutdensity}
as follows%
\begin{equation} \label{jumpdensityv}
\rho(\chi)|_{\chi=\tanh{(v/2)}}=1-\frac{  c^2}{  v}+O(v^{-2}), \quad v\gg 1.
\end{equation}

For $v\gg 1$, one obtains from equation \e{zetadef} that $1-\chi\ll 1$ and $v=-\ln\left ( \frac{1-\chi}{2} \right )+O(1-\chi).$ Then the density \e{jumpdensityv}
takes the following form
\begin{equation} \label{jumpdensityvasymp}
\rho(\chi)=1+\frac{  c^2}{  \ln\left ( \frac{1-\chi}{2} \right )}+O\left (\frac{1}{\ln^2\left ( {1-\chi}{} \right )}\right ).
\end{equation}

Equation \e{jumpdensityvasymp} implies a unit value
\begin{equation} \label{rho1}
\rho(1)=1
\end{equation}
and the divergence of the derivative
\begin{equation} \label{rho1der}
\frac{d\rho(\chi)}{d\chi}\simeq\frac{  c^2}{  (1-\chi)\ln^2\left ( \frac{1-\chi}{2} \right )} \to \infty \ \text{for} \ \chi\to 1.
\end{equation}

\section{Numerical procedure to analyze the structure of sheets of Riemann surface for Stokes wave by ODE integration}
\label{sec:Numericalstructuresheets}

We use Pad\'e approximants of Stokes wave found in Part I
\citep{DyachenkoLushnikovKorotkevichPartIStudApplMath2016} and
provided both in tables of Part I, through the electronic attachment
to Ref. \citet{DyachenkoLushnikovKorotkevichPartIArxiv2015} and at
the web link~\citet{PadePolesList} in the following form
\begin{equation}
\label{approx_cut}  z_{pade}(u)\equiv u+\I  y_b+ \sum\limits_{n =
1}^{N} \dfrac{\gamma_n}{\tan(u/2) - \I \chi_n},
\end{equation}
with the numerical values of $y_b$, the pole positions $\chi_n$ and
the complex residues $\gamma_n$ ($n=1,\ldots, N $) given there.
These data of Pad\'e approximation allow to recover the Stokes wave
at the real axis $w=u$ (and similar at $\zeta=Re(\zeta)$  in the
complex  $\zeta$-plane) with the relative accuracy of at least
$10^{-26}$ (for the vast majority of numerical cases the actual
accuracy is even higher by  several orders of magnitude).

Analytical continuation of the Pad\'e approximant \e{approx_cut}
from $u$ to $w\in \mathbb{C}$  is given by the straightforward
replacing of $u$ by $w$. That analytical continuation is accurate
for $w\in \mathbb{C}^-$ but looses precision for $w\in \mathbb{C}^+$
in the neighbourhood of  the branch cut $w\in[\I v_c,\I\infty)$
where the discrete sum  \e{approx_cut} fails to approximate the
continuous paramterization \e{branchcutdensity} of the branch cut.
Thus a significant loss of precision compare to $10^{-26}$ occurs
only if the distance from the  given value of  $\zeta$ to the branch cut is smaller or
comparable with the distance between neighbouring  values of
$\chi_n$ in equation  \e{approx_cut}.

Numerical integrations
of ODE \e{Stokeszuzubarexplicit} (and occasionally ODE \e{Stokeszuzubarzetaexplicit}) in this Section were performed using 9(8)th order explicit Runge-Kutta algorithm with adaptive stepping embedded into Mathematica 10.2 software. That algorithm is the implementation of Ref. \citet{VernerNumericalAlgorithms2010}  and is based on the embedded pair of 9th and 8th order methods  with higher order method used for  the  adaptive step-size control. We used the numerical precision of 55 digits and reached the accuracy $10^{-30}$ to make sure that no significant accumulation of ODE integration error occurs in comparison with $10^{-26}$ precision of the Pad\'e approximants of Stokes waves. We also independently verified the accuracy of the numerical ODE integration by comparing with the analytical results of Section \ref{sec:SeriesexpansionsStructureRiemannsurface} in the neighborhoods of $\zeta=\pm\I\chi_c$  in multiple sheets of Riemann surface.

At the first step  of our investigation, ODE \e{Stokeszuzubarexplicit} was solved numerically to find the approximation $z_{ODE}(\zeta)$ for $z(\zeta)$ with $\zeta\in\mathbb{C}^+$
in the first and the second sheet of Riemann surface  using the approximants   \e{approx_cut} for  $\bar z$ and $\bar z_\zeta$.  Here the first (physical) sheet of Riemann surface corresponds to $z(\zeta)$ with fluid occupying $\zeta\in \mathbb{C}^-.$   The second (non-physical) sheet is reached when the branch cut $\zeta\in [\I \chi_c, \I]$ (or equivalently $w\in[\I v_c,\I\infty)$) is crossed from the first sheet.  That ODE was solved with initial conditions at real line $\zeta=Re(\zeta)$  by integrating along different contours in $\zeta\in\mathbb{C}^+$.    A high precision of at least $10^{-30}$ was achieved in ODE solver to avoid any significant additional loss of precision compare with  $10^{-26}$  precision of  equation \e{approx_cut}.
That ODE solution used    $\bar z_{pade}$ and $(\bar z_{pade})_\zeta$ which through the complex conjugation corresponds to the approximants  \e{approx_cut} in $\zeta\in\mathbb{C}^-$ thus avoiding any loss of precision compare with  $10^{-26}$. We stress here that the use of Pad\'e approximation is the auxiliary tool  which does not make any difference in the final result because it matches the precision of Fourier series. The Fourier series of Part I can be used directly instead of Pad\'e approximats which however would require significant increase of computational resources to reach the same precision.

\begin{figure}
\includegraphics[width=1.06\textwidth]{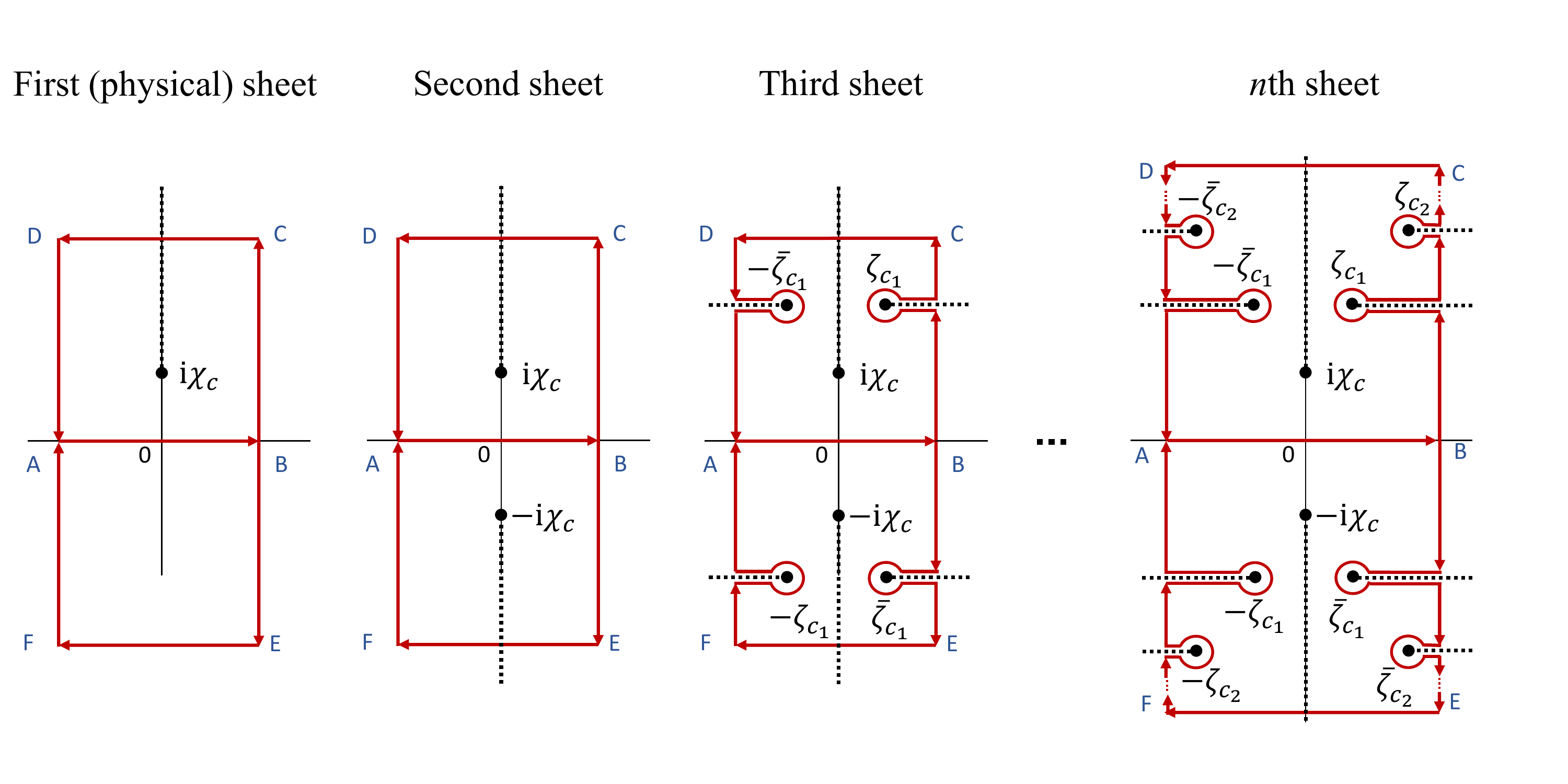}
\vspace{-0.3cm}
 \caption{  A schematic of integrating contours in different sheets of  Riemann surface  in the complex variable $\zeta$  \e{zetadef} near the origin. The first (physical) sheet has a square root singularity only at $\zeta=\I\chi_c$ in $\mathbb{C}^+.  $  Then integrating ODE  \e{Stokeszuzubarexplicit} over the closed contour $ABCDA$  provides the analytical continuation into the second sheet of Riemann surface as the branch cut (dashed line) is crossed.  As a result, $z(\zeta)$ does not return to its initial value at the origin $0$. In contrast,  integrating   ODE  \e{Stokeszuzubarexplicit} over the closed contour $ABEFA$ (or over  $ABCDA$  provided its height falls below $\zeta=\I\chi_c$), one does not cross the branch cut so  $z(\zeta)$  returns to the same value at the origin with $z(\zeta)$ remaining in the first sheet. The second sheet has the second square root branch point singularity at $\zeta=-\I\chi_c$  at the lower
 complex half-plane $\mathbb{C}^-$. Integrating over contour $ABEFA$ in the second sheet results in the analytical continuation of $z(\zeta)$ into the third sheet of Riemann surface.  Starting from the third sheet, extra square root branch points appear away from the  imaginary axis.  Branch cuts for these off-axis singularities are chosen to be extended horizontally as shown by dashed lines in the two right panels. The number of these branch points grows with the growth of the sheet number as schematically shown in the right panel. We avoid crossing these branch by modifying contours $ABCDA$ and $ABEFA$ as shown in the two right panels. Note that   these two contours must by symmetric with respect to  the real line  even if the chosen pair of off-axis singularities (symmetric with respect to the imaginary axis) are located only in one of the complex half-planes $\mathbb{C}^+$ and $\mathbb{C}^-$ in the given sheet. This is because  $\bar z(\zeta)$ is needed for the integration of ODE   \e{Stokeszuzubarexplicit}.         }
 \label{fig:contour1}
\end{figure}

\begin{figure}
(a)\includegraphics[width=0.46\textwidth]{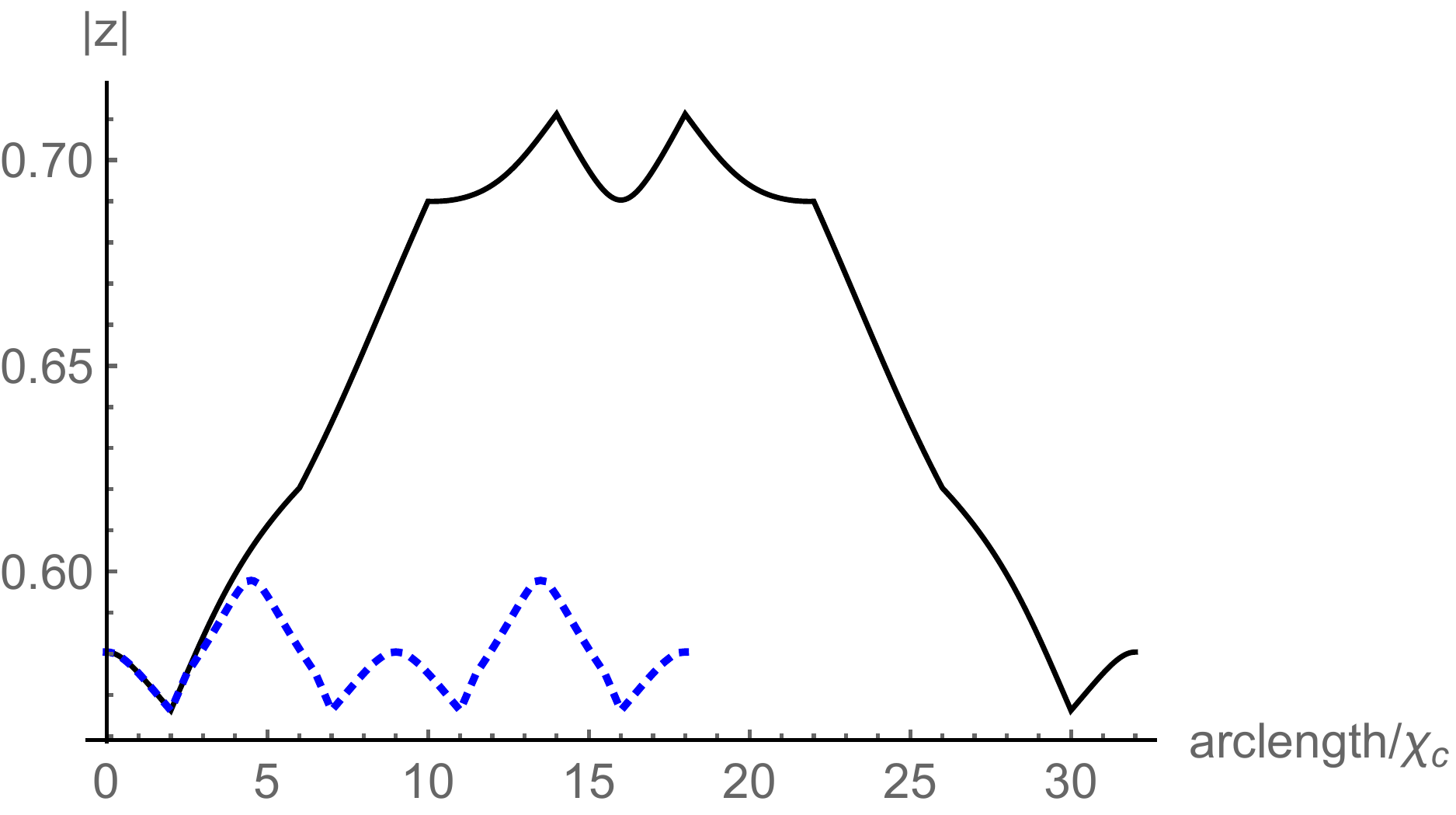}
(b)\includegraphics[width=0.46\textwidth]{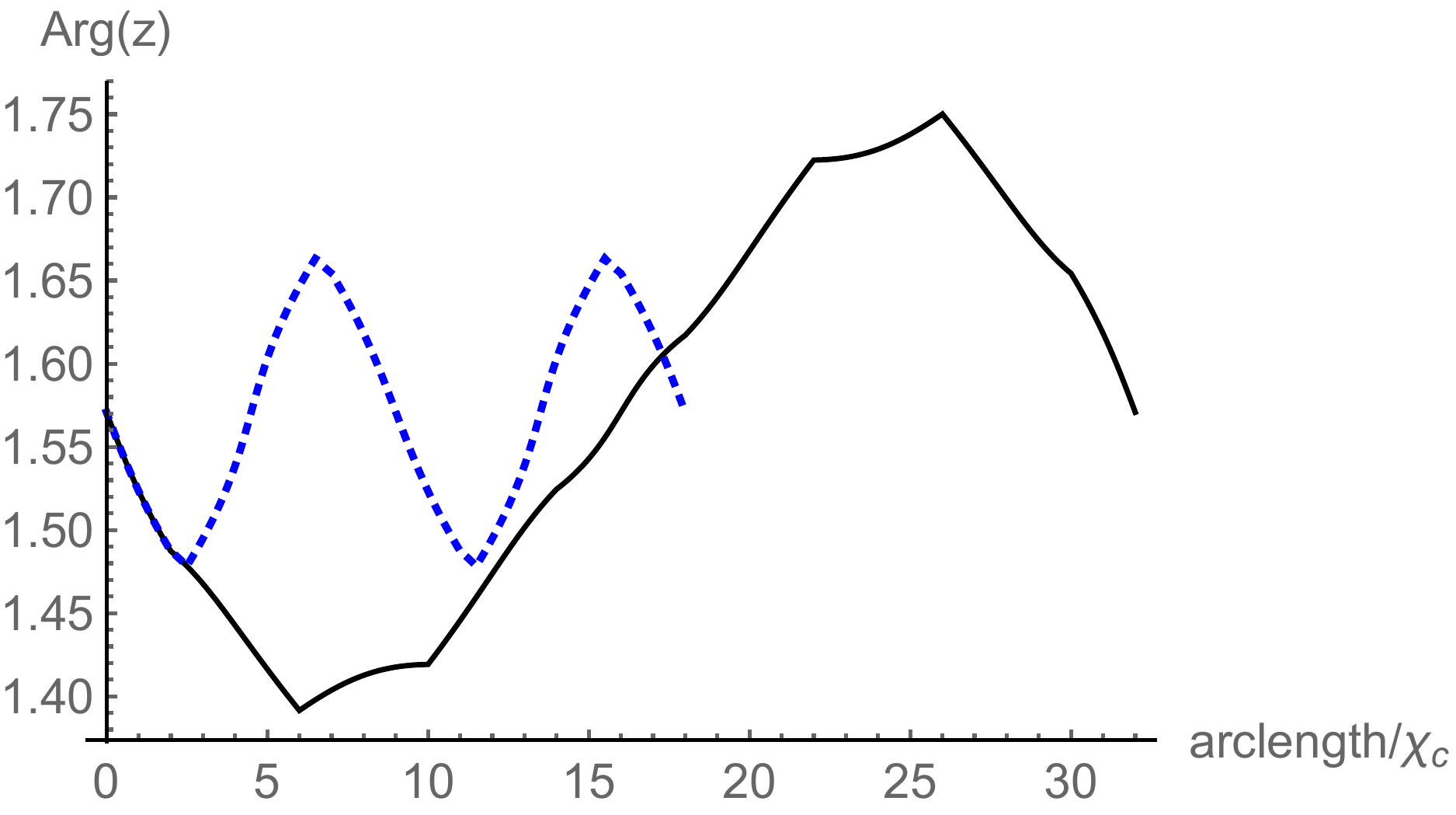}
\vspace{-0.01cm}
 \caption{  The amplitude $|\tilde z(\zeta)|$ (a) and the argument $\text{Arg}(\tilde z(\zeta))$  (b) vs. arclength in the variable  $\zeta$ (scaled by $\chi_c$) along the closed contour $0BCDA0$ shown on the left panel of Figure \ref{fig:contour1}  (the contour is passed twice in the counterclockwise direction) for ODE integration (provides the analytical continuation of Stokes wave in the complex plane) of  Stokes wave solution  with $H/\lambda=0.1387112446\ldots$ (corresponds to $\chi_c=3.0056373876\ldots\cdot10^{-3}$, see Table \ref{tab:H.chic} of Appendix
\ref{sec:TablesStokesWaveschic} for details on numerical Stokes wave).  The contour width is $|AB|=2\,|0B|=2\chi_c$. Solid lines are for the contour height   $|BC|=4\chi_c$  (the contour  $0BCDA0$  twice crosses the branch cut $[\I\chi_c,\I]   $ in counterclockwise direction) and dotted lines are for  the contour height   $|BC|=\chi_c/2$   (for that height the contour  $0BCDA0$  does not crosses the branch cut $[\I\chi_c,\I]   $ as well as the total arclength is smaller). It is seen that  solid lines are periodic over the total  arclength (two round trips around the contour  $0BCDA0$  are needed to return to the initial value $z(0)$ in the first sheet of Riemann surface) compare with  the half-arclength periodicity of dotted lines (the contour   $0BCDA0$  is located in the first sheet only with one roundtrip sufficient to return to the initial value $z(0)$).          }
 \label{fig:contoursheet1p2}
\end{figure}

The left panel of Figure \ref{fig:contour1} shows a typical $0BCDA0$
rectangular contour for ODE integration  which was used for the
analytical continuation of Stokes wave into the second sheet of
Riemann surface.  The ODE solution in the second sheet is obtained
when integrating contour crosses  the branch cut $[\I\chi_c,\I].$
The second subsequent crossing of that branch cut returns
$z_{ODE}(\zeta)$ to the first sheet confirming the square root branch
point at $\zeta=\I \chi_c$. Figure \ref{fig:contoursheet1p2}
provides a numerical example of such double crossing.  In other
words, it was found that ODE integration along any closed contour in
$\zeta\in\mathbb{C}^+$ with double crossing of the branch cut (twice
integrating along  $0BCDA0$ ) always returns the solution to  the
original one.  If the  height of  $0BCDA0$  contour is made   smaller
than $\chi_c$ then there is no crossing of the branch cut and
$0BCDA0$ integration returns to the initial value after a single
round trip as shown by dashed curves of Figure
\ref{fig:contoursheet1p2}. In similar way, if the height of $0BCDA0$
exceeds 1 then there is no crossing and   $z_{ODE}(\zeta)$ stays in
the first sheet (crossing of the branch cut $(\I,\I\infty)$
corresponds to the jump on $2\pi$ in $u$ direction in $w$ plane while there is no jump in $\tilde z$ because of $2\pi$-periodicity).

We also verified that there are no singularities in the limit $|Re(\zeta)|\to\infty$ by switching to ODE integration  \e{Stokeszuzubarzetaexplicit} in $w$ variable. In that limit  $Re(w)\to\pm \pi$ which allows  to extend the contour in $w$  over the entire $2\pi$ period in $u$ direction (in $\zeta$ variable it would require to integrate over the infinite interval $-\infty< Re(\zeta)<\infty)$. For all subsequent cases in this section it is assumed that such integration in $w$ was performed to check the limit  $Re(w)\to\pm \pi$.

The second step of our investigation was to find $z(\zeta)$  by integrating  ODE \e{Stokeszuzubarexplicit} in the second sheet with $\zeta\in\mathbb{C}^-$ using the complex conjugate of  $z_{ODE}(\zeta), \ \zeta\in\mathbb{C}^+ \ $ found at previous step to approximate  $\bar z$ and $ \bar z_\zeta$ .  Initial condition at that step was at the real line  $\zeta=Re(\zeta)$  with  $z(\zeta)$   obtained at the step one for the second sheet.

    The second step reveals a new square root singularity at $\zeta=-\I\chi_c$ in the second sheet. Similar to the step one, the double integration over the contour
     $ABEFA$ shows that $z(\zeta)$ returns to its original value confirming  that  $\zeta=-\I\chi_c$    is the square root branch point. Crossing of the branch cut  $[-\I\chi_c,-\I]$
     (corresponds to that new branch point  $\zeta=-\I\chi_c$) allows to go into the third sheet of Riemann surface. At that crossing one has to simultaneously cross from the first to the second sheets for  $\bar z$ and $\bar z_\zeta$ which again are the complex conjugate of  $z_{ODE}(\zeta), \ \zeta\in\mathbb{C}^+ \ $ found at previous step.  It was found that the third sheet has branch points both at  $\zeta=\I\chi_c$   and  $\zeta=-\I\chi_c.$ In a similar way to previous steps, at the step three one crosses  the branch cut  $[\I\chi_c,\I]$ to go into the fourth sheet of Riemann surface which found to has branch points both at  $\zeta=\I\chi_c$   and  $\zeta=-\I\chi_c.$  At the step four one crosses  the branch cut  $[-\I\chi_c,-\I]$ to go into the firth sheet of Riemann surface which again has branch points both at
     $\zeta=\I\chi_c$   and  $\zeta=-\I\chi_c$ etc. At each sheet, $\bar z, \bar z_\zeta$ used in integration of   ODE \e{Stokeszuzubarexplicit} is behind by one in sheet number
     to the current sheet, i.e. values of  $\bar z, \bar z_\zeta$ from the first, second, third etc. sheets are used for ODE integration in the second, third, fourth etc. sheets, respectively.
After exploring  several hundreds of sheets for different values of $\chi_c$, one concludes  that the number of sheets is
infinite. The double integration over the contours $ABCDA$    and  $ABEFA$ shows that   $\zeta=\I\chi_c$   and  $\zeta=-\I\chi_c$
are square root branch points in all sheets.  In the next section this conjecture is  strengthened by the analysis of expansions
at $\zeta=\pm \I \chi_c$ in these multiple sheets.

 Starting from the third sheet, extra square root branch points appear away from the  imaginary axis. The existence of these  off-axis singularities are closely related  to the analysis of Section \ref{sec:conjecturetwothirds}. They are located significantly more far away from the origin  than the on-axis $\zeta=\pm\I\chi_c$  singularities. These singularities appear at each sheet starting from the third one in pairs located symmetrically with respect to the imaginary axis as schematically shown in Figure \ref{fig:contour1}.    The symmetric location of pairs of singularities are required  from the symmetry condition

\begin{equation} \label{Stokessymm}
\bar z(-\zeta)=-z(\zeta).
\end{equation}
That symmetry condition results from  the symmetry  $y(x)=y(-x)$  of
Stokes wave in physical variables. The location of the first pair of
off-axis square root singularities  at $\zeta=\bar
\zeta_{c_1}$ and $\zeta=- \zeta_{c_1}$ is schematically shown in the
third panel of Figure \ref{fig:contour1}. By adaptively increasing
 the horizontal and vertical sizes of  the contour
$ABEFA$ of Figure \ref{fig:contour1}, we found that for $\chi_c\ll 1$
the first pair of off-axis square root singularities are located in
the third sheet at $\zeta=\bar \zeta_{c_1}$ and $\zeta=-
\zeta_{c_1}$ with
\begin{equation} \label{zetac1}
\bar \zeta_{c_1}\simeq(17.1719 - \I \,10.7734)\chi_c.
\end{equation}
Other off-axis pairs are located even more far away both from the
real and imaginary axes as schematically shown in Figure
\ref{fig:contour1}. Branch cuts for all off-axis singularities are
chosen to be extended horizontally as shown by dashed lines in the
two right panels of  Figure \ref{fig:contour1}.  For
$|A0|,\,|0B|\lesssim 17.1719 \chi_c,$  one can use the same contour
as in the left panel for all sheets. However, for larger values of
 $|A0|,\,|0B|$ one has to bypass off-diagonal singularities
as shown in the two right panels of  Figure \ref{fig:contour1} to
keep the enumeration of the sheets as described above (based on
on-axis $\zeta=\pm\I\chi_c$  singularities).     The number of
off-axis branch points grows with the increase of the sheet number.
We also performed double integration over closed contours around multiple off-axis singularities and found that each of them is the square root branch point.

By-product of ODE integration of this section is that one can also calculate the jump $-2\pi\rho(\chi)$  (see  equation \e{branchcutdensity}) at the branch cut of the first sheet with the high precision. E.g. one can start ODE integration at $\zeta=0$ in the first sheet and integrate until reaching a small neighborhood of $\zeta=\I\chi_c$ without crossing the branch cut    $\zeta\in[\I\chi_c,\I].$ After that one can integrate ODE independently along  two line segments     $\zeta= [\pm\epsilon+\I\chi_c,\pm\epsilon+\I]$, $\epsilon\to 0$ and calculate a difference between these two integrations recovering $\rho(\chi)$ with the precision of our simulations $\sim 10^{-26}.$ A comparison of that high precision  $\rho(\chi)$ with the numerical approximation of
$\rho(\chi)$ obtained in Part I from the continuous limit of Pad\'e approximation (see Figs.  6b, 7 and 8 in Part I) confirmed the numerical error order estimates of Section 4.2 of Part I. Also we found that equations   \e{zjump}-\e{rho1der} are also in the excellent agreement with the numerical values of   $\rho(\chi)$ confirming the asymptotical analysis of Section \ref{sec:AsymptoticStokeswave}.

ODE of the type \e{Stokeszuzubarexplicit} was  numerically
integrated  in Ref. \citet{TanveerProcRoySoc1991} based on the
Taylor series  representation of Stokes wave in the physical sheet
(the additional conformal mapping from the unit disk used in Ref.
\citet{TanveerProcRoySoc1991} into the half-plane $\mathbb{C}^-$ of
$\zeta$ \e{zetadef} makes that Taylor series similar to the Fourier
series representation of Part I
\citep{DyachenkoLushnikovKorotkevichPartIStudApplMath2016}). That
representation allowed Ref.  \citet{TanveerProcRoySoc1991} for the
first time to extend the numerical integration into the upper half
$\mathbb{C}^+$ of the second Riemann sheet and demonstrate the
existence of the square root branch point there.  Thus a  numerical ODE integration of  Ref.
\citet{TanveerProcRoySoc1991}  is similar
to our first step of this Section restricted to $\mathbb{C}^+$ only.

Note that it was assumed throughout   this Section  that any crossing by the ODE integration contour of both   $[\I,\I\infty] $  and    $[-\I\infty,-\I]$ is avoided. Such crossing would be harmless  in the first sheet because of $2\upi$-periodicity of $\tilde z(w)$. However,   starting from the second sheet, $\tilde z(w)$ is generally non-periodic in $w.  $ Thus the branch cuts    $[\I,\I\infty] $  and    $[-\I\infty,-\I]$  cannot be ignored any more contrary to the case of the first sheet case discussed in the Introduction.
It implies that a crossing of these branch cut provides the additional sheets of Riemann surface. We however do not explore these sheets here because they have the distance $1$ from the real axis in $\zeta$ plane for any  value of $\chi_c$ thus  not contributing to the formation of the  limiting Stokes  wave.

\section{Series expansions at $\zeta=\pm i\chi_c$ and structure of Riemann surface for Stokes wave}
\label{sec:SeriesexpansionsStructureRiemannsurface}

Equation \e{Stokeszuzubar} together with the definition \e{bardef}
shows that singularities at  $\zeta=\pm i\chi_c$ are coupled through
complex conjugation. We found in Part I
\citep{DyachenkoLushnikovKorotkevichPartIStudApplMath2016} that
there is only one singularity (a square root branch point) in the
first (physical) sheet of Riemann surface which corresponds to the
finite complex $w$ plane. In addition, there is a singularity at
$\zeta=\I $ which is the complex infinity $w=\I \infty$ and is
discussed in Section \ref{sec:AsymptoticStokeswave}. Following Part
I we chose the line segment $[\I \chi_c,\I]$ as the branch cut
connecting these two singularities in the first sheet of Riemann
surface as sketched on the left panel of Figure
\ref{fig:sheets_sketch}. Singularity at $\zeta=-\I\chi_c$ is not
allowed in the first sheet because $z $ is analytic in the fluid
domain $w\in \mathbb{C}^-.$

Consider the expansions in $l$th sheet of Riemann surface
\begin{align} \label{zetachicserp}
z_{l,+}(\zeta)=\sum\limits_{j=0}^\infty \I e^{ \I j\pi/4}a_{+,l,j}(\zeta- \I \chi_c)^{j/2}, \quad l=1,2,\ldots,
\end{align}
and
\begin{align} \label{zetachicserm}
z_{l,-}(\zeta)=\sum\limits_{j=0}^\infty  \I e^{- \I j\pi/4}a_{-,l,j}(\zeta+  \I \chi_c)^{j/2},\quad l=1,2,\ldots,
\end{align}
where subscripts $``+"$ and $``-"$ mean expansions at
$\zeta=\I \chi_c$ and $\zeta=-\I \chi_c$, respectively. Here the branch
cuts of $(\zeta- \I \chi_c)^{1/2}$ and $(\zeta+ \I \chi_c)^{1/2}$
are assumed to extend from $\zeta= \I \chi_c$ upwards and from
$\zeta=- \I \chi_c$ downwards, respectively as shown in Figure
\ref{fig:sheets_sketch}. Often a location of the branch cut of
square root is taken  on the negative real axis of the argument. To
use that standard  agreement about a location of the branch cut, one
can replace  $(\zeta- \I \chi_c)^{j/2}$ and $(\zeta+ \I
\chi_c)^{j/2}$ in  equations \e{zetachicserp} and \e{zetachicserm}
by $(-\I)^{j/2}( \I \zeta+\chi_c)^{j/2}$ and   $\I^{j/2}(- \I
\zeta+\chi_c)^{j/2}$, respectively.

Following Section \ref{sec:Numericalstructuresheets}, we enumerate sheets of Riemann surface according to the branch
points $\zeta=\pm\I \chi_c$ as follows. A crossing of the branch cut
$[\I\chi_c,\I]$ in the counterclockwise direction means going from
$l=2n-1$th sheet of Riemann surface to $l=2n$th sheet with
$n=1,2,\ldots$. Case $l=1$ corresponds to the physical sheet of
Riemann surface. Similarly, crossing of a branch cut $[-\I\chi_c,-\I]$
in the counterclockwise direction means going from $l=2n$th sheet of
Riemann surface to $l=2n+1$ sheet with $n=1,2,\ldots$. Plugging
expansions \e{zetachicserp} and \e{zetachicserm} into equation
\e{Stokeszuzubar} and collecting terms of the same order of
$(\zeta\pm \I\chi_c)^{j/2}$ result in the following relations
\begin{equation}\label{amser}
\begin{split}
& a_{-,2n,1} =0, \\
&a_{-,2n,2} = \frac{-2}{{1-\chi_c}^2}, \\
&a_{-,2n,3}=\frac{16 c^2}{3 \left(1-\chi_c^2\right)^2 a_{+,2n-1,1} (c^2-{a}_{+,2n-1,0}-{a}_{-,2n,0})}, \\
&a_{-,2n,4}= \frac{2\chi_c}{{(1-\chi_c}^2)^2}+ \frac{4 c^2}{{(1-\chi_c}^2)^2 (c^2 - a_{+,2n-1, 0} - a_{-,2n, 0})^2}\\
&\qquad\quad - \frac{8 c^2 [2 + (-1 + \chi_c^2) a_{+,2n-1, 2}]}{{(1-\chi_c}^2)^3a_{+,2n-1, 1}^2 (c^2 - a_{+,2n-1, 0} - a_{-,2n, 0})},
\\
&\ldots
\end{split}
\end{equation}
for $n\ge 1$ and
\begin{equation}\label{apser}
\begin{split}
&a_{+,2n+1,1} =-\frac{16 {c^2}}{3 \left(1-{\chi_c}^2\right)^2 a_{-,2n,3} ({c^2}{-a}_{-,2n,0}+{a_{+
,2n+1,0}})},  \\
&a_{+,2n+1,2} = \frac{2}{{1-\chi_c}^2}+ \frac{128 c^4}{{9(1-\chi_c}^2)^4a_{-,2n, 3}^2 (c^2 - a_{-,2n, 0} - a_{+,2n+1, 0})^3}\\
&\qquad\qquad+ \frac{32 c^2 [-2\chi_c + (1 - \chi_c^2)^2 a_{-,2n, 4}]}{{9(1-\chi_c}^2)^4a_{-,2n, 3}^2 (c^2 - a_{-,2n, 0} - a_{+,2n+1, 0})},
\\
&a_{+,2n+1,3} =\ldots, \\
&\ldots\\
\end{split}
\end{equation}
for $n\ge 1$.

One cannot take  $n=0$ in  equation   \e{apser} which corresponds to $l=1$ (the physical sheet of Riemann surface). This special case has to be considered separately because in the physical sheet there is no singularity at $\zeta=-\I \chi_c$ (no singularity inside fluid domain). It implies that
\begin{equation} \label{aml}
a_{-,1,2j+1}=0 \quad \text{for} \quad j=0,1,2,\ldots
\end{equation}
Solving equations \e{zetachicserp}, \e{zetachicserm} and \e{Stokeszuzubar}
for $l=1$ with the series expansion at $\zeta=-i\chi_c$ subject to the condition \e{aml}
results in the following expressions
\begin{equation}\label{ap1ser}
\begin{split}
&a_{+,1,0} =c^2-a_{-,1,0} , \\
&a_{+,1,1} =\frac{-2^{3/2} {c}}{\left(1-{\chi_c}^2\right)^{1/2}\left[  (2+(1-\chi_c^2)a_{-,1,2})\right ]^{1/2}},  \\
&a_{+,1,2} =\frac{4}{3 \left(1-{\chi_c}^2\right)}-\frac{a_{-,1,2}}{3},  \\
&a_{+,1,3} =-\frac{\left[2+(1-\chi_c^2)a_{-,1,2})\right ]^{5/2}}{2^{1/2}18c\left(1-{\chi_c}^2\right)^{3/2}}
\\&\qquad\quad +\frac{2^{1/2}c\left[2\chi_c -
  2 \chi_c (-1 + \chi_c^2) a_{-, 1, 2} + (-1 + \chi_c^2)^2 a_{-, 1, 4}\right ]}
{\left(1-{\chi_c}^2\right)^{3/2}\left[2+(1-\chi_c^2)a_{-,1,2})\right ]^{3/2}},
\\
&a_{+,1,4} =\ldots, \\
&\ldots\\
\end{split}
\end{equation}

Expressions \e{ap1ser} are uniquely determined by values of $c$,
$\chi_c$ and  $a_{-,1,2j}$, $j=0,1,2,\ldots$, where all expressions
under square roots are positive and the principle branch  of all
square roots is assumed. In contrast, the expressions     \e{amser} and
\e{apser}  are not the unique solutions of equations
\e{zetachicserp}, \e{zetachicserm} and \e{Stokeszuzubar}. In
addition to the solution  \e{amser}, one can obtain two more
spurious solutions for  $a_{-,2n,j}$.  However, these spurious
solutions do not correspond to  Stokes wave. One spurious solution
has  $a_{-,2n,2j+1}=0 \quad \text{for} \quad j=1,2,\ldots $, i.e. it
does not have a singularity at $\zeta=-i\chi_c$. The second  spurious
solution  has either a radius of convergence well below $\chi_c$ or
even the zero radius of convergence. Both solutions  are spurious
because they cannot have the same value  in the region of overlap
of the disks of convergence of both expansions   \e{zetachicserp}
and \e{zetachicserm}. After spurious solutions for    $a_{-,2n,j}$
are discarded, one obtains the unique solution   \e{amser} as well as the solution    \e{apser}   $a_{+,2n+1,j}$  also
turns to be uniquely defined.  Another peculiar property of the
solution \e{amser} is that  $a_{-,2n,1} =0$ while   $a_{+,2n+1,1}\ne
0$ as given by the solution    \e{apser}.

R.h.s. of  equation \e{apser} provides the explicit expressions for
the coefficients $a_{+,2n+1,j}$, $j=1,2,,\ldots$, for $2n+1$th sheet
of Riemann surface   through  the coefficients $a_{-,2n,j_1}$, $0\le
j_1\le j+2$ at $2n$th sheet. The only coefficient which remains
unknown is the zeroth coefficient $a_{+,2n+1,0}$ for each $n\ge 1$.
In a similar way, r.h.s. of  equation \e{amser} provides the
explicit expressions for the coefficients $a_{-,2n,j}$
$j=1,2,,\ldots$, for $2n$th sheet of Riemann surface    through the
coefficients $a_{+,2n-1,j_1}$, $0\le j_1\le j-2,$ $j=1,2,,\ldots$,
at $2n-1$th sheet. The only coefficient which remains unknown is the
zeroth coefficient $a_{-,2n,0}$ for each $n\ge 1$.

The explicit  expressions for $a_{+,2n+1,j}$ and $a_{-,2n,j}$ turn
cumbersome with the increase of $j$ beyond values shown explicitly
in  equations   \e{amser} and  \e{apser}.  The  explicit expression
$a_{+,2n+1,j}$ and $a_{-,2n,j}$ were obtained with the help of
symbolic computations in Mathematica 10.2 software. These expressions
were used to calculate values of all  coefficients   $a_{+,2n+1,j}$
and $a_{-,2n,j}$ for $j\ge 1$ numerically with any desired precision
(typically we used quadruple  (quad) precision with 32 digits
accuracy and took into account all $j$ in the range $1\le j\le
200$). The remaining coefficients $a_{-,2n,0}$ and $a_{+,2n+1,0}$
for each $n\ge 1$ as well as the numerical value of $\chi_c$ were
determined by a numerical procedure which is described below in
Sections \ref{sec:chicmatching} and \ref{sec:apmmatching}.

Values of $a_{+,2n+2,j}$ and $a_{-,2n+1,j}$ are obtained from $a_{+,2n+1,j}$, and $a_{-,2n,j}$ by the following relations
\begin{equation}\label{aplusaminusrelations}
\begin{split}
& a_{+,2n+2,j}= (-1)^{j}a_{+,2n+1,j}, \quad n=0,1,2,\ldots,  \\
& a_{-,2n+1,j}= (-1)^{j}a_{-,2n,j}, \quad n=1,2,\ldots,
\end{split}
\end{equation}
which immediately follows from the condition at the crossing of
branch cuts.

\subsection{Finding of  $\chi_c,$ from matching the series expansions at $\zeta=\pm\I   \chi_c$ in the first sheet }
\label{sec:chicmatching}

Equations
 \e{ap1ser}   determine  values of  $a_{+,1,j}$, $j=0,1,2,\ldots$ from  $a_{-,1,2j}$, $j=0,1,2,\ldots$ thus relating the
 series expansions at $\zeta= -\I \chi_c$ and   $\zeta= \I \chi_c$ at the first sheet. The series at  $\zeta= -\I \chi_c$ is given by equation
 \e{zetachicserm} with $l=1$ together with the condition \e{aml}. That series contains only integer powers of $\zeta+\I\chi_c$. The disk of
 convergence $|\zeta+\I\chi_c|<r $ of that series is determined by the branch point at $\zeta=\I\chi_c $ which implies that the radius of convergence is  $r=2\chi_c$.
  The series at  $\zeta= \I \chi_c$ at the first sheet is given by equations
 \e{zetachicserp},  \e{ap1ser} and contains both integer and half-integer powers of  $\zeta-\I\chi_c$.  The disk of convergence $|\zeta-\I\chi_c|<r $ is determined by the branch point at  $\zeta= -\I \chi_c$  of the second sheet. Thus the radius of convergence is also  $r=2\chi_c$.
In other words, the radius of convergence of the series \e{zetachicserp},   \e{ap1ser} in the physical sheet is determined by the singularity in the second (non-physical sheet).

Numerical values of the coefficients   $a_{-,1,2j}$, $j=0,1,2,\ldots$ are immediately obtained by the differentiation of the Pad\'e approximants of Part I for each numerical value of $H/\lambda$. Accuracy of that approximation of the coefficients   $a_{-,1,2j}$ is checked by plugging  these numerical values into the series \e{zetachicserm} with $l=1$ and using \e{aml}.  For numerical evaluation that series is truncated   into a finite sum

\begin{align} \label{zetachicsermfinite}
z_{{1,-,sum}}(\zeta)=\sum\limits_{j=0}^{j_{max}}  \I e^{- \I j\pi/4}a_{-,1,j}(\zeta+  \I \chi_c)^{j/2}=\sum\limits_{j=0}^{j_{max}/2}  \I e^{- \I j\pi/2}a_{-,1,2j}(\zeta+  \I \chi_c)^{j},
\end{align}
where $j_{max}$ is chosen sufficiently large to match the numerical
precision of Pad\'e approximants. It is convenient  to evaluate that
sum at  $\zeta=0$ which is well inside the disk of convergence
$|\zeta+\I\chi_c|<2\chi_c$. It was found that $j_{max}=200$ at
$\zeta=0$ is well sufficient to reach a numerical precision about
quad precision $\sim10^{-32}$ of simulations of Part I. That
numerical value of $j_{max}$ (sufficient to reach  quad precision)
is only weakly dependent on $H/\lambda$. To understand that weak
dependence one can note that $|\zeta+\I\chi_c|_{\zeta=0}=\chi_c$
which is   one-half of the radius of convergence of the series
\e{zetachicserm}. The asymptotics of the terms of the series
\e{zetachicserm} for large $j$ is determined by the radius of
convergence as follows $|a_{-,1,2j}/a_{-,1,2j+2}|\simeq 2\chi_c.  $
Then the truncation of the series  \e{zetachicserm}  by the finite
sum \e{zetachicsermfinite} with $j_{max}=200$ gives the error
 $\sim a_{-,1,j_{max}}\chi_c^{j_{max}/2}\sim  2^{-j_{max}/2}\sim10^{-30 }$ in comparison with  Pad\'e approximation of Stokes wave at $\zeta=0.$

It worth to note here that the number of derivatives $j_{max}/2=100$ which was reliably recovered above from  Pad\'e approximation is really large which demonstrates the highly superior  efficiency  of Pad\'e approximation compare with Fourier series.  E.g., if instead Pad\'e approximation of Stokes wave, one uses the Fourier series representation of Stokes wave, then the number of derivatives calculated   from that series with a high numerical precision would be limited to    just a few (about 10-20 derivatives if  the relative error $\sim 1$ in derivatives is allowed).

To obtain numerical values of   $a_{+,1,j}$, $j=0,1,2,\ldots $ from equations
 \e{ap1ser}   one also has to know the numerical value of $\chi_c$.
Part I described a numerical procedure to recover $\chi_c$
 with the accuracy $\sim 10^{-10}$ which is significantly below the accuracy $\lesssim 10^{-26}$ of numerical Stokes solution itself and its Pad\'e approximation. In this paper to greatly improve that precision of $\chi_c,$ one sets a condition that $\chi_c$ is chosen in such a way to allow the series \e{zetachicserp} to recover the value of $z(0)$ with the accuracy  better than $\sim 10^{-26}.$

Similar to equation \e{zetachicsermfinite}, the series \e{zetachicserp} is truncated   to a finite sum
\begin{align} \label{zetachicserpfinite}
z_{{1,+,sum}}(\zeta)=\sum\limits_{j=0}^{j_{max}}  \I e^{ \I j\pi/4}a_{+,1,j}(\zeta-\I \chi_c)^{j/2},\quad
\end{align}
where we again choose that $j_{max}=200$ which is well enough to
match quad precision $\sim10^{-32}$. Contrary to equation
\e{zetachicsermfinite}, the sum  \e{zetachicserpfinite} includes
also half-integer powers of $\zeta-\I \chi_c$ because
$\zeta=\I\chi_c$ is the square root branch point. Using equations
\e{ap1ser} and  \e{zetachicserpfinite} with the numerical values of
$a_{-,1,2j}$, $j=0,1,2,\ldots,$ obtained as described in the
beginning of this Section, one finds in the first sheet a numerical
value of $ z_{{1,+,sum}}(0)$ for each numerical value of $\chi_c$.
Then numerical Newton (secant) iterations are performed over
$\chi_c$ aiming to ensure that $  z_{{1,+,sum}}(0)$ converges to
$\simeq z_{pade}(0)$, i.e. $\chi_c$ is chosen such that  $
z_{{1,+,sum}}(0)$  recovers  the value of $z_{pade}(0). $
 It provides $\chi_c$ with the precision at least $10^{-26}$ which is
limited by the precision of  Pad\'e approximation. Part I also
demonstrated the calculation of Stokes wave well beyond  quad
precision by using variable precision arithmetics with the achieved
accuracy $\sim 200$ digits thus increasing of accuracy for  $\chi_c$
is also  possible if needed. Table \ref{tab:H.chic} of Appendix
\ref{sec:TablesStokesWaveschic} provides numerical values of
$\chi_c$ which correspond to  Pad\'e approximations  of Stokes wave
found in Part I.

\subsection{Finding of
$a_{+,2n+1,0}$ and $a_{-,2n,0}$ from matching the series expansions
at $\zeta=\pm\I  \chi_c$  in the second, third etc. sheets}
\label{sec:apmmatching}

The procedure for finding numerical values of $\chi_c$ and    $a_{+,1,j}$, $j=0,1,2,\ldots$ described in Section \ref{sec:chicmatching}, together with equations  \e{amser}
and \e{aplusaminusrelations}, allows to immediately find $a_{-,2,j},$  $j=1,2,\ldots$ for each given value of  $a_{-,2,0}$. Similar to equations \e{zetachicsermfinite} and \e{zetachicserpfinite}, a notation is used such that    $  z_{{l,+,sum}}(\zeta)$ and     $  z_{{l,-,sum}}(\zeta)$ are the finite sums corresponding to the truncation of the series     $  z_{{l,+}}(\zeta)$  \e{zetachicserp}  and     $  z_{{l,-}}(\zeta)$   \e{zetachicserm}, respectively. We assume that  $j_{max}\simeq 200$ for  these finite sums in the sheets $l=1,2,\ldots.$
 The numerical Newton iterations at the first step are performed over  $a_{-,2,0}$ aiming to ensure that    $  z_{{2,-,sum}}(0)$      converges to $z_{{2,+,sum}}(0)$. At the second step, the Newton iterations  allow to find   $a_{+,3,0}$ by matching $z_{{3,-,sum}}(0)$ and $z_{{3,-,sum}}(0).     $ In a similar way, the third, fourth etc. steps allow to find  $a_{-,4,0}$,  $a_{+,5,0}$,  $a_{-,6,0}$, $a_{+,7,0},  \ldots$   Then using equations  \e{aplusaminusrelations}, one obtains values of $a_{+,n,0}$ and $a_{-,n,0}$ for all positive integer $n$ completing the analytical continuation of Stokes wave into the disks $|\zeta\pm\I \chi_c|<2\chi_c$ in the infinite number of sheets of Riemann surface.

The result of that analytical continuation was compared with the
analytical continuation by ODE integration of Section
\ref{sec:Numericalstructuresheets} giving the excellent agreement  which
is only limited by the standard numerical accuracy $\sim 10^{-26}$
of Stokes wave in the physical sheet. Increasing that accuracy of
analytical continuation is straightforward by increasing $j_{max}$
for the finite sums     $  z_{{l,+,sum}}(\zeta)$ and     $
z_{{l,-,sum}}(\zeta)$   (and similar by increasing the accuracy for
ODE integration) provided Stokes wave precision is increased. Table
\ref{tab:asheet} of Appendix \ref{sec:TablesStokesWaveschic}
provides a sample of numerical values  of   $a_{-,2n,0}$,
$a_{+,2n+1,0},$ for $n=1,2,3$  obtained by the numerical method outlined in
this Section.

\section{Singularities of Stokes wave for finite values of $w$ }
\label{sec:squarerootsingularities}

  \citet{MalcolmGrantJFM1973LimitingStokes} and \citet{TanveerProcRoySoc1991}   showed that the only possible singularity in the finite complex upper half-plane of the physical sheet of Riemann surface is of square root type. This result is consistent
both with   the simulations of Part I
\citep{DyachenkoLushnikovKorotkevichPartIStudApplMath2016} and
numerical integration  of   ODE \e{Stokeszuzubarexplicit} in Section
\ref{sec:Numericalstructuresheets}.

The analysis of  \citet{TanveerProcRoySoc1991} is based on a version of  equation \e{Stokeszuzubarzeta} together with the assumption of the analyticity of $\bar z(w)$ in $\mathbb{C}^+$ for the fist sheet of Riemann surface. Assume that one performs ODE integration in the second, third etc. sheets of Riemann surface as described in Sections  \ref{sec:Numericalstructuresheets} and \ref{sec:SeriesexpansionsStructureRiemannsurface} with $z(w)$ at the $n$th sheet coupled to $\bar z(w)$ in the $n-1$th sheet. Here the counting of sheets follows Section \ref{sec:Numericalstructuresheets} and assumes that  $-\pi<Re(w)<\pi$, $|Im(w)|<\infty$ for all sheets.
   Then the analysis of   \citet{TanveerProcRoySoc1991} can be immediately generalized  to the $n$th sheet at values of $w=w_1$ such that $z(w)$ has no  singularity at $w=\bar w_1 $ in the $n-1$th sheet (see equation  \e{Stokeszuzubarzetaexplicitqpsi3} below). Coupling of square root singularities at $\zeta=\pm\I\chi_c$ which is studied in Section  \ref{sec:SeriesexpansionsStructureRiemannsurface}  however goes beyond the analysis of   \citet{TanveerProcRoySoc1991}.

The series expansions of  Section
\ref{sec:SeriesexpansionsStructureRiemannsurface} show that square
root singularities can occur at any finite values of $w=w_1$ away
from the real axis. It was found in Section
\ref{sec:SeriesexpansionsStructureRiemannsurface} that each square
root singularity  can either have a sister square root singularity
at the complex conjugated point  $w=\bar w_1$ in the same sheet  or
can exists without the sister singularity at $w=\bar w_1$ thus going
beyond the case analyzed by \citet{TanveerProcRoySoc1991}. A
question still remains if any other type (beyond square root) of
coupled singularities at $w=w_1$ and $w=\bar w_1$ in the same sheet
is possible.

Going from the first sheet to the second one, then from the second one to the third one etc., one concludes
that the only way for the singularity other than square root to
appear is to be coupled with the square root singularity in the
previous sheet. Otherwise if would violate the above mentioned generalization of the
result of  \citet{TanveerProcRoySoc1991} to the arbitrary sheet.  Consider a general  power law
singularity of $z(w)$ at $w=w_1$ coupled with the square root
singularity of $\bar z (w)$ at $w=\bar w_1$. We write that general
singularity in terms of double series as follows
\begin{equation} \label{wbarwsingularities}
z(w)=\sum\limits_{n,m}c_{n,m}(w-w_1)^{n/2+m\alpha},
\end{equation}
where $\alpha$ is the real constant, $c_{n,m}$ are complex constants and $n,m$ are integers.  By shifting $n$ and $m$ one concludes that without loss of generality one can assume that %
\begin{equation} \label{alpharange}
0<\alpha<1/2.
\end{equation}
After the
complex conjugation, the square root singularity of $ z(w)$  at
$w=\bar w_1$ is given by the following series
\begin{equation} \label{wbarwsingularities2}
\bar z(w)=\sum\limits_{n=0}^\infty d_n(w-w_1)^{n/2},
\end{equation}
where $d_n$ are the complex constants. Coupling of $z(w)$   and $\bar z(w)$  in equation  \e{Stokeszuzubarzetaexplicit} explains why half-integer powers $n/2$ must be taken into account in equation  \e{wbarwsingularities}.
It is convenient to transform from $w$ into a new  complex variable
\begin{equation} \label{qdef}
q\equiv (w-w_1)^{1/2}.
\end{equation}
Equation \e{Stokeszuzubarzetaexplicit} for the new variable $q$ takes the following form
\begin{equation} \label{Stokeszuzubarzetaexplicitq}
z_q=\frac{4q^2c^2}{\bar z_q[\I(z-\bar z)+c^2]}.
\end{equation}
The series  \e{wbarwsingularities} is then transformed into

\begin{equation} \label{wbarwsingularities2q}
z(q)=\sum\limits_{n,m}c_{n,m}q^{n+2m\alpha},
\end{equation}
while the series \e{wbarwsingularities2}
 runs over integer powers, %
\begin{equation} \label{wbarwsingularitiesq}
\bar z(q)=\sum\limits_{n=0}^\infty d_nq^{n}.
\end{equation}
The series  \e{wbarwsingularities2q} can be also called by $\psi$-series, see e.g. \citet{Hille1997}.
If $\alpha$ is the rational number then in equation \e{wbarwsingularities}   one can gather together all terms with the same power of $q$ thus reducing equation \e{wbarwsingularities}   to Puiseux series
\begin{equation} \label{wbarwsingularities2p}
z(q)=\sum\limits_{n=-\infty}^\infty \tilde c_nq^{2n/k},
\end{equation}
where $k$ is the positive integer.

If one additionally restricts that there is no essential singularity at $q=0$ then one has to replace \e{wbarwsingularities2q}  with the truncated series

\begin{equation} \label{wbarwsingularities2qtruncated}
z(q)=\sum\limits_{n\ge n_0,m\ge m_0}c_{n,m}q^{n+2m\alpha},
\end{equation}
for the integer constants $n_0$ and $m_0.$ Plugging in equations \e{wbarwsingularitiesq} and  \e{wbarwsingularities2qtruncated}
into the Stokes wave equation \e{Stokeszuzubarzetaexplicitq}, moving the denominator to the l.h.s. in equation  \e{Stokeszuzubarzetaexplicitq}    and collecting terms with the same power of $q$, starting from the lowest power, one obtains that $2\alpha$ must be integer   for any values of $n_0$ and $m_0$ and all values of $d_n.$   Thus no new solutions in the form   \e{wbarwsingularities2qtruncated} exists beyond what was found in Section  \ref{sec:SeriesexpansionsStructureRiemannsurface}.

One can also study singularities using the classification of movable and fixed singularities in nonlinear ODEs of 1st order in the general form $z_q=f(q,z)$ (see e.g \cite{Golubev1950,Ince1956,Hille1997}). Position of fixed singularities for the independent complex variable $q$ is determined by the properties of ODE, i.e. by singularities of the function $f(q,z)$. In contrast, the position of movable singularity is not fixed but typically is determined by an arbitrary complex constant. To analyze singularities, it is convenient to introduce a new unknown

\begin{equation} \label{xidef}
\xi(q)\equiv\frac{1}{\I(z-\bar z)+c^2}.
\end{equation}
Then equation \e{Stokeszuzubarzetaexplicitq} takes the following form

\begin{equation} \label{Stokeszuzubarzetaexplicitqxi}
\xi_q=\I\bar z_q\xi^2-\frac{4\I q^2c^2\xi^3}{\bar z_q},
\end{equation}
where one reminds that $\bar z(q)$ is assumed to be known and is determined by $z(\bar q)$ from the previous sheet of Riemann surface. Equation  \e{Stokeszuzubarzetaexplicitqxi} has a cubic polynomial r.h.s in $\xi$ which ensures that it has a movable square root singularity
\begin{equation} \label{movablexi}
\xi=\sum\limits_{n=-1}^\infty c_n (q-C)^{n/2},
\end{equation}
provided $C\ne 0$, $\bar z_q(C)\neq 0$ (see e.g \cite{Golubev1950,Ince1956,Hille1997}), where $c_n$ and $C$ are the complex constants.
Using equations \e{qdef}, \e{xidef} and the condition  $C\ne 0$, one recovers the expansion  \e{wbarwsingularities2} with $w_1$ replaced by $w_1+C^2$ thus the movable singularity \e{movablexi} is reduced to the square root singularity in $w$.

Equation    \e{Stokeszuzubarzetaexplicitqxi}  has a fixed singularity at $q=0$ provided $\bar z_q(0)\neq 0.$ To show that one  uses a new unknown $\psi\equiv1/\xi$ to transform equation  \e{Stokeszuzubarzetaexplicitqxi} into

\begin{equation} \label{Stokeszuzubarzetaexplicitqpsi}
\psi_ q=\frac{-\I\bar  z^2_q\psi+4\I q^2c^2}{\bar z_q\psi },
\end{equation}
which has $0/0$ singularity in r.h.s. for $q=\psi=0$   satisfying the criteria for the existence of fixed singularity (see \cite{Golubev1950,Hille1997}).

Consider now a particular case  $\bar z_q(0)= 0$ and  $\bar z_{qq}(0)\ne 0 $ which corresponds to the expansion \e{amser}. One can define a new function
\begin{equation} \label{gfunc}
g(q)\equiv  \frac{\tilde z_q}{q}, \quad g(0)\ne 0
\end{equation}
and rewrite equation \e{Stokeszuzubarzetaexplicitqpsi}  as follows
\begin{equation} \label{Stokeszuzubarzetaexplicitqpsi2}
\psi_ q=\frac{-\I q g(q)^2\psi+4\I q c^2}{g(q)\psi },
\end{equation}
Generally this equation still has  a fixed singularity because of $0/0$ singularity in r.h.s.. However, in a particular case  when $g(q)$ is the even function of $q$, i.e. one can define the  function $\tilde g(q^2)\equiv g(q)$ which is analytic in the variable $\tilde q\equiv q^2$ at $q=0$. This case means that $z(w)$ is analytic at $w=w_1.$ Then one transforms equation \e{Stokeszuzubarzetaexplicitqpsi2} into
the equation
\begin{equation} \label{Stokeszuzubarzetaexplicitqpsi3}
\psi_ {\tilde q}=\frac{-\I\tilde  g(\tilde q)^2\psi+4\I c^2}{ 2\tilde g(\tilde q)\psi },
\end{equation}
which does not have a fixed singularity. Equation  \e{Stokeszuzubarzetaexplicitqpsi3} together with equation \e{movablexi} reproduces the result of \citet{TanveerProcRoySoc1991}  applied to all sheets of Riemann surface.

Thus the approaches  reviewed in Refs. \cite{Golubev1950,Ince1956,Hille1997}
applied to equation \e{Stokeszuzubarzetaexplicitq} are consistent with square root singularities and the series expansions of Section   \ref{sec:SeriesexpansionsStructureRiemannsurface} for all sheets of Riemann surface. However, these approaches cannot exclude the possibility of existence of other types of singularity. Note that  examples given in   \cite{Golubev1950,Ince1956,Hille1997}
also show that the existence of the fixed singularity in ODE at the point $q=0$ does not necessary mean that the singularity occurs  in the general ODE solution $z(q)$ at that point.

One concludes that a coupling of the essential singularity at $w=w_1$ with the square root singularity at $w=\bar w_1$ cannot be
excluded neither by the series analysis used in equations    \e{qdef}-\e{wbarwsingularities2qtruncated}
 nor by looking at the fixed ODE singularities  through equations \e{xidef}-\e{Stokeszuzubarzetaexplicitqpsi2}. However, the simulations of Section \ref{sec:Numericalstructuresheets} and series expansions of Section  \ref{sec:SeriesexpansionsStructureRiemannsurface} clearly indicates the absence of any singularities beyond square roots  in all sheets of Riemann surface for non-limiting Stokes wave in
 $\mathbb{\zeta\in \mathbb{C}} \diagdown((-\I\infty,-\I]\cup ([\I,\I\infty))$ (i.e. everywhere in the  complex plane $\mathbb{C}$ except the branch cuts $(-\I\infty,-\I]$ and $[\I,\I\infty)$). In the first sheet the branch cuts   $(-\I\infty,-\I]$ and $[\I,\I\infty)$ are not significant as explained in the Introduction, and  the only non-square root singularity exists at $\zeta=\I$, see Section \ref{sec:AsymptoticStokeswave}. It is conjectured here that non-square root singularities do not appear in all sheets of Riemann surface for
 $\mathbb{\zeta\in \mathbb{C}} \diagdown((-\I\infty,-\I]\cup ([\I,\I\infty)).
$

As discussed at the end of Section \ref{sec:Numericalstructuresheets},  singularities are possible at the boundary of the strip $Re(w)=\pm \pi$ which corresponds to the  branch cuts     $[\I,\I\infty) $  and    $[-\I\infty,-\I) $ in $\zeta$ plane. However, these branch cuts are separated by the distance $\pi$ from the origin in $w$ plane (or by the distance $1$ in $\zeta$ plane) and they cannot explain the formation of the limiting Stokes wave as $v_c\to 0.$ The same is true for the singularity at $w\to \I\infty$ ($\zeta\to \I$)  analyzed in Section \ref{sec:AsymptoticStokeswave}.

\section{Conjecture on recovering of $2/3$ power law of limiting Stokes wave from infinite number of nested square root singularities of non-limiting Stokes wave as $\chi_c\to 0$  }
\label{sec:conjecturetwothirds}

One concludes from Sections
\ref{sec:SeriesexpansionsStructureRiemannsurface}  and
\ref{sec:squarerootsingularities} that the only possibility for the
formation of $2/3$ power law singularity \e{w2p3} of the limiting
Stokes wave is through the merging together the infinite number of
square root singularities from different sheets of Riemann surface
in the limit $v_c\to 0$. The total number of square root
singularities could be either finite or infinite for $v_c>0$, both
cases are compatible with the expansions of Section
\ref{sec:SeriesexpansionsStructureRiemannsurface} (although the
infinite number of singularities appears to hold  for the generic
values of the expansion coefficients of Section
\ref{sec:SeriesexpansionsStructureRiemannsurface}). Both numerical
ODE integration of Section \ref{sec:Numericalstructuresheets} and
series expansions of Section
\ref{sec:SeriesexpansionsStructureRiemannsurface}  reveal that the
number of sheets of Riemann surface related to singularities at
$\zeta=\pm\I \chi_c$ exceeds several hundreds for a wide range of
numerical values  $10^{-7}\lesssim\chi_c\lesssim 0.2.$ It suggests
that the number of  sheets is infinite for all values of $v_c.$ In
any case, the number of singularities must be infinite as $v_c\to
0.$

Here  a conjectured  is made that the limiting Stokes wave occurs as the limit  $v_c\to 0$ of the following leading order solution

\begin{align} \label{2p3square}
&z=\I\frac{c^2}{2}+c_1\chi_c^{1/6}\sqrt{\zeta-\I\chi_c}\nonumber \\&+\frac{(3c)^{2/3}}{2}e^{-\I\pi/6}\left[(\zeta-\I\chi_c)^{1/2}+(-2\I\chi_c)^{1/2} \right ]
\sqrt{\alpha_1\chi_c^{1/4}+\sqrt{(\zeta-\I\chi_c)^{1/2}+(-2\I\chi_c)^{1/2}} } \nonumber\\
&\times\sqrt{\alpha_3\chi_c^{1/16}+\sqrt{\alpha_2\chi_c^{1/8}+ \sqrt{\alpha_1\chi_c^{1/4}+\sqrt{(\zeta-\I\chi_c)^{1/2}+(-2\I\chi_c)^{1/2}} }}} \nonumber \\
&\times\sqrt{\alpha_{2n+1}\chi_c^{1/2^{2n+2}}+\sqrt{\alpha_{2n}\chi_c^{1/2^{2n+1}}+\sqrt{\ldots+ \sqrt{\alpha_1\chi_c^{1/4}+\sqrt{(\zeta-\I\chi_c)^{1/2}+(-2\I\chi_c)^{1/2}} }}}} \nonumber \\
&\times\cdots+\frac{(3c)^{2/3}}{2}e^{-\I\pi/6}\left[(\zeta-\I\chi_c)^{1/2}+(-2\I\chi_c)^{1/2} \right ]
\sqrt{\tilde\alpha_1\chi_c^{1/4}+\sqrt{(\zeta-\I\chi_c)^{1/2}+(-2\I\chi_c)^{1/2}} }\text{}
\nonumber \\
&\times\sqrt{\tilde\alpha_3\chi_c^{1/16}+\sqrt{\tilde\alpha_2\chi_c^{1/8}+ \sqrt{\tilde\alpha_1\chi_c^{1/4}+\sqrt{(\zeta-\I\chi_c)^{1/2}+(-2\I\chi_c)^{1/2}} }}} \nonumber \\
&\times\sqrt{\tilde\alpha_{2n+1}\chi_c^{1/2^{2n+2}}+\sqrt{\tilde\alpha_{2n}\chi_c^{1/2^{2n+1}}+\sqrt{\ldots+ \sqrt{\tilde\alpha_1\chi_c^{1/4}+\sqrt{(\zeta-\I\chi_c)^{1/2}+(-2\I\chi_c)^{1/2}} }}}} \nonumber \\
&\times\cdots\text{+h.o.t.},
\end{align}
which is the infinite product of increasingly nested square roots.
This conjecture was first  presented at the conference talk
\citet{LushnikovIMACS2015}.  Equation \e{2p3square}  has two terms
with  nested  roots, one with  nonzero complex constants
$\alpha_1,\, \alpha_2,\, \alpha_3,\, \alpha_4,\ldots$ and another
with  nonzero complex constants  $\tilde \alpha_1,\, \tilde
\alpha_2,\, \tilde \alpha_3,\, \tilde \alpha_4,\ldots$ which are
related through complex conjugation as follows
\begin{equation} \label{alphabars}
\tilde \alpha_1=\bar \alpha_1 e^{-\I\pi/4}, \ \tilde \alpha_2=\bar \alpha_1 e^{-\I\pi/8}, \ldots, \ \tilde \alpha_n=\bar \alpha_n e^{-\I\pi/2^{n+1}}, \ldots
\end{equation}
All these constants including another complex constant $c_1$ are of the order $O(1)$ independent of $\chi_c$. The relations \e{alphabars} ensures that the symmetry condition
\e{Stokessymm} is satisfied.

At
 $\zeta\gg \chi_c$ one obtains from equation \e{2p3square} using the asymptotic of  products of all square roots  that
\begin{equation} \label{2p3sum}
z\propto \zeta^{1/2+1/8+1/32+1/128+\dots}=\zeta^{2/3}
\end{equation}
exactly reproducing the Stokes solution  \e{w2p3} while the term $c_1\chi_c^{1/6}\sqrt{\zeta-\I\chi_c}$ vanishes as $\chi_c\to 0.$ For small but finite $\chi_c$, the limiting Stokes solution   \e{w2p3} is valid for $\chi_c\ll \zeta\ll 1$, as seen from equation  \e{2p3square}. For $\zeta\sim 1$, the higher order terms denoted by $\text{h.o.t.}$ both in equations  \e{w2p3} and  \e{2p3square}, becomes important such as the term with the irrational power

\begin{equation} \label{mudef1}
\propto \zeta^\mu, \ \mu=1.4693457\ldots
\end{equation}
 \citep{MalcolmGrantJFM1973LimitingStokes,Williams1981}.

Different branches of all nested square roots in equation  \e{2p3square} choose different sheets of Riemann surface following the numeration of sheets used in Section
 \ref{sec:Numericalstructuresheets}.
In particular, the principal branch of $(\zeta-\I\chi_c)^{1/2}$ in the expression $g(\zeta)\equiv(\zeta-\I\chi_c)^{1/2}+(-2\I\chi_c)^{1/2}$  corresponds to the first sheet. To understand that one expands $g(\zeta)$ at $\zeta=-\I \chi_c$ which results in
\begin{equation} \label{gplus}
g_{+}(\zeta) =2(-2\I\chi_c)^{1/2}+\frac{\zeta+\I\chi_c}{2(-2\I \chi_c)^{1/2}}+O(\zeta+\I\chi_c)^2,
\end{equation}
where the subscript ``+" means taking the principle branch of  $(\zeta-\I\chi_c)^{1/2}$. For the second (negative) branch of $(\zeta-\I\chi_c)^{1/2}$  one obtains that %
\begin{equation} \label{gminus}
g_{-}(\zeta) =-\frac{\zeta+\I\chi_c}{2(-2\I \chi_c)^{1/2}}+O(\zeta+\I\chi_c)^2
\end{equation}
with the subscript $``-"$ meaning that second branch.

The expression $g(\zeta)$ enters under the most inner square root into each term of the product in equation  \e{2p3square}. Then using the expansion \e{gplus} one obtains that the series expansion of equation  \e{2p3square} at   $\zeta=-\I \chi_c$ contains only nonnegative integer powers of $\zeta+\I\chi_c$ thus confirming that $z(\zeta)$ is analytic at  $\zeta=-\I \chi_c$. It means that the condition \e{aml} is satisfied. In contrast, taking  the expansion \e{gminus} one obtains that the series expansion of equation  \e{2p3square} at   $\zeta=-\I \chi_c$ is of the type \e{zetachicserm} containing  nonnegative half-integer powers of $\zeta+\I\chi_c $ as expected for all sheets starting from the second sheet.  In addition, the term $g(\zeta)$ in the square brackets in  equation  \e{2p3square} ensures that $a_{-,2n,1} =0$  as required by equation \e{amser}.
The expansion of equation   \e{2p3square}  at  $\zeta=\I \chi_c$   has half-integer powers of  $\zeta-\I\chi_c $   for both branches of $g(\zeta)$ thus being consistent with the square root singularity  at $\zeta=\I \chi_c$ in all sheets of Riemann surface including the first sheet in agreement with equations \e{zetachicserp}, \e{apser} and \e{ap1ser}.

Choosing two possible branches of all other nested square roots (besides
the most inner square root) in equation   \e{2p3square} one obtains
the expansions  \e{zetachicserp} and  \e{zetachicserm}, at
$\zeta=\pm\I \chi_c$        with different coefficients $a_{+,l,j}$
and $a_{-,l,j}$ at each $l$th sheet. The values of these
coefficients are determined by both values of  $\chi_c, \, c_1, \,
\alpha_1,\, \alpha_2,\, \alpha_3,\, \alpha_4,\ldots$ together with
the contribution from $\text{h.o.t}$ terms in equation
\e{2p3square}. One concludes that the ansatz    \e{2p3square} is
consistent with the properties of non-limiting Stokes wave studied
in this paper which motivates the conjecture   \e{2p3square}. Also
coefficients $\alpha_1,\, \alpha_2,\, \alpha_3,\ldots $ determine
additional square root branch points which are
  located away from the imaginary axis at distance larger than $\chi_c$  from the origin in the third and higher sheets of Riemann surface. Values of these coefficients can be determined from the locations of branch points thus independently recovered from ODE integration similar to described in Section  \ref{sec:Numericalstructuresheets} (see also Section \ref{sec:conjecturenumerical} for the example of recovering of $\alpha_1$).

\subsection{Numerical verification of the conjecture}
\label{sec:conjecturenumerical}

Now we provide a numerical demonstration of the efficiency of the conjecture   \e{2p3square} by using the simplest nontrivial approximation of equation \e{2p3square} which takes into account only the three-fold nested roots as follows

\begin{align} \label{2p3squaretruncated}
&z\simeq\left (\I\frac{c^2}{2}+c_0\chi_c^{2/3}\right )+c_1\chi_c^{1/6}\sqrt{\zeta-\I\chi_c}+c_2\zeta\nonumber \\&+c_3\chi_c^{1/24}\left[(\zeta-\I\chi_c)^{1/2}+(-2\I\chi_c)^{1/2} \right ]
\sqrt{\alpha_1\chi_c^{1/4}+\sqrt{(\zeta-\I\chi_c)^{1/2}+(-2\I\chi_c)^{1/2}} }
\nonumber \\&+\bar c_3e^{\frac{-3\I\pi}{8}}\chi_c^{1/24}\left[(\zeta-\I\chi_c)^{1/2}+(-2\I\chi_c)^{1/2} \right ]
\sqrt{\bar \alpha_1e^{\frac{-\I\pi}{4}}\chi_c^{1/4}+\sqrt{(\zeta-\I\chi_c)^{1/2}+(-2\I\chi_c)^{1/2}} }
\end{align}
where we added the term $c_2\zeta$  as well the constants $c_0,$ $c_2$ and $c_3$ to approximate the neglecting of other nested roots (which includes $\alpha_2, \alpha_3, \ldots$) in comparison with equation  \e{2p3square}. In other words,  we approximate all more than three-fold nested roots in equation \e{2p3square}  by   Taylor series expansion keeping only the constant and linear terms in $\zeta $ in that expansion. The constant terms result in adding the constants $c_3$ and $\bar c_3$ in front of nested roots as well as in the small correction $c_0\chi_c^{2/3}$ to the constant term   $\I c^2/2$  of equation \e{2p3square}). The linear  terms result in the appearance of the constant $c_2$  replacing higher order nested roots of  equation  \e{2p3square}.  Also the factor $\bar c_3e^{\frac{-3\I\pi}{8}}$ ensures the symmetry   \e{Stokessymm} and the additional scaling $\chi_c^{1/24}$ provides for $\zeta\sim\chi_c$ the same total scaling $\propto \chi_c^{2/3}$   of the nested square roots of equation \e{2p3squaretruncated}  as in equation  \e{2p3square}.
The approximation \e{2p3square} can be valid only up to moderately large values of $Re(\zeta)=\zeta$ (a comparison with equation  \e{2p3square}
suggests that it could be valid up to values of $\zeta$ in about several tens of $\chi_c$ after that higher order nested roots must come into play).

To find the numerical values of $\alpha_1$, $c_0,$ $c_1, \ c_2$ and
$c_3$ without any fit we use the following procedure. At the first
step we determine  from the contour integration of Section
\ref{sec:Numericalstructuresheets} the location  \e{zetac1} of the
first pair of off-axis square root singularities (located at
$\zeta=\bar \zeta_{c_1}$ and $\zeta=- \zeta_{c_1}$).
Then we solve   equation $\alpha_1\chi_c^{1/4}+\sqrt{(\bar \zeta_{c_1}-\I\chi_c)^{1/2}+(-2\I\chi_c)^{1/2}}=0$ (corresponds to the zero under three-fold square root in equation  \e{2p3square}) in the third sheet of the Riemann surface which together with  \e{zetac1} gives that %
\begin{equation} \label{alpha1num}
\alpha_1\simeq-0.0955383 - \I \,1.8351.
\end{equation}
Note that the second square root $\zeta=-\zeta_{c_1}$ (symmetric with respect to the imaginary axis)
is ensured by the similar term with $\bar \alpha_1e^{\frac{-\I\pi}{4}}$  in equation  \e{2p3squaretruncated}. At the  second step we expand equation \e{2p3square} in the first sheet of Riemann surface in powers of $(\zeta-\I\chi_c)^{1/2}$. After that we match the first five coefficients of that expansion to the analytical expressions of the coefficients $a_{+,1,j}$, $j=1,\ldots,5$, of the   expansion \e{ap1ser} obtained in Section \ref{sec:SeriesexpansionsStructureRiemannsurface}. That matching results in the explicit expressions for $c_0,$ $c_1, \ c_2$ and $c_3$. E.g., for Stokes wave with $\chi_c=2.9691220994\ldots\cdot 10^{-7}$ (corresponds to  the last line of  table \ref{tab:H.chic} of Appendix \ref{sec:TablesStokesWaveschic}, see that Appendix for more details on the numerical Stokes wave) we obtain that  $c_0=\I\,17.1920\ldots,$ $c_1=e^{-\I\pi/4}3.81499\ldots, \ c_2=-1.8779\ldots$ and $c_3=1.42696\ldots-\I\, 1.8849\ldots$ Changing of $\chi_c$ by several orders of magnitude results in changing these coefficients only within the range $5-10\%.$

\begin{figure}
\includegraphics[width=0.94\textwidth]{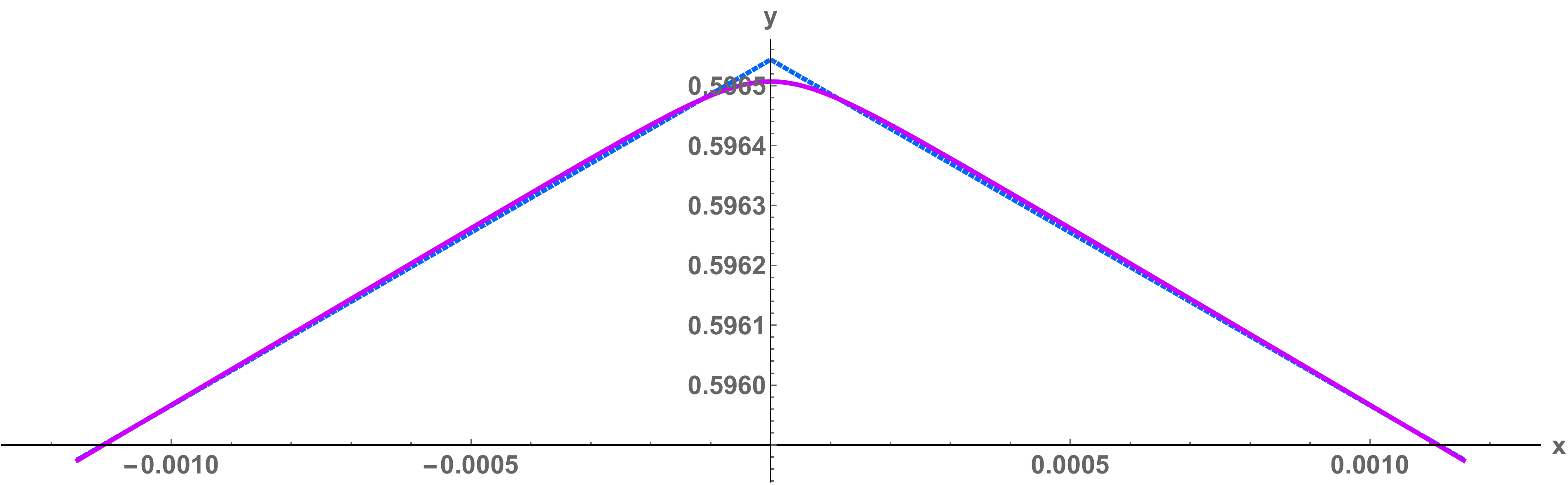}
\vspace{-0.01cm}
 \caption{  A comparison of the limiting Stokes wave (dotted line) with  equation \e{2p3squaretruncated} and the numerical solution for Stokes wave (last two of these are shown by the single solid line because they are visually indistinguishable with maximum difference between them $\simeq  4\cdot10^{-6}$) for  $\chi_c=2.9691220994\ldots\cdot 10^{-7}$. Solid line corresponds to $-50\chi_c\le\zeta \le 50\chi_c.$      }
 \label{fig:nested_limiting}
\end{figure}

We demonstrated the efficiency of the obtained numerical values of
$c_0,$ $c_1, \ c_2$ and $c_3$  in two independent ways. In the first
way, it was checked that the coefficients $a_{+,1,j}$ of equation
\e{ap1ser} for $j=6,\ldots$ are well reproduced (within $4\%$ and
$7\%$ accuracy for $j\le 10$ and $j\le 100$, respectively) by the
expansion of equation \e{2p3squaretruncated} with the same numerical
values of  $c_0,$ $c_1, \ c_2$ and $c_3$. It implies that  the
approximate expression  \e{2p3squaretruncated} captures the
significant property of the convergence of the series  \e{ap1ser}
rather than being just the match of a few first terms of that
series. The second way of efficiency demonstration  is provided in
Figure \ref{fig:nested_limiting}, where the excellent agreement is
shown between the numerical solution of Stokes wave from Part I
\citep{DyachenkoLushnikovKorotkevichPartIStudApplMath2016} and the
expression  \e{2p3squaretruncated} for $-50\chi_c\le \zeta\le
50\chi_c$. That range of $\zeta$ is  far beyond the disk of
convergence $|\zeta-\I\chi_c|<2\chi_c$ of the series   \e{ap1ser}.

\section{Concluding remarks}
\label{sec:Conclusion}

In summary, it was found that the Riemann surface corresponding to non-limiting Stokes wave consists of the infinite number of sheets  corresponding to the  infinite number of square root branch points  located at   $\zeta=\pm \I \chi_c$ in all sheets except the first sheet. The first (physical) sheet has only one singularity at   $\zeta=\I \chi_c$ while avoiding singularity at $\zeta=-\I \chi_c$ which ensures that Stokes wave represents the analytical solution inside the fluid domain. Two ways of analytical continuation into all these sheets were used, the first one is based on ODE integration of Section  \ref{sec:Numericalstructuresheets} and the second one is based on the coupled series expansions  \e{zetachicserp}-\e{ap1ser} in half-integer powers at    $\zeta=\pm \I \chi_c$.

To go beyond the disks of convergence $|\zeta\pm\I \chi_c|<2\chi_c$ of the series expansions \e{zetachicserp}-\e{ap1ser},  it is conjectured in Section \ref{sec:conjecturetwothirds}  that the leading order form of non-limiting Stokes wave consists of the infinite number of nested square roots  \e{2p3square}. These nested square roots  can recover the series expansions \e{zetachicserp}-\e{ap1ser} within their disks of convergence $|\zeta\pm\I\chi_c|<2\chi_c.$ For $|\zeta\pm\I\chi_c|\gg \chi_c$, well beyond these disks of convergence, the asymptotic \e{2p3sum} is valid thus ensuring that the nested square roots  form $2/3$ power law singularity of the limiting Stokes wave in the limit $\chi_c\to 0$.

There are two other infinite sequences of Riemann sheets resulting from (a) off-axis square root singularities  in the third and higher sheets of Riemann surface as analyzed in Sections   \ref{sec:Numericalstructuresheets} and  \ref{sec:conjecturetwothirds}, and (b) the singularity at $\zeta=\I$ (corresponds to $w=\I\infty$) which involves logarithms as analyzed in Section \ref{sec:AsymptoticStokeswave}. However, these extra sheets do not contribute to  the qualitative change of power law singularity from $1/2$ (non-limiting Stokes wave) to $2/3$ (limiting Stokes wave) near the origin as given by the asymptotic \e{2p3sum}.  However, these extra sheets are expected to  be important for the analysis of Stokes wave for  $\zeta\sim 1$, where the higher order terms becomes important such as the term \e{mudef1} with the irrational power
\citep{MalcolmGrantJFM1973LimitingStokes,Williams1981}.
The analysis of these terms is  beyond the scope of this paper.
 These terms might be also essential to answer the question  left open by Refs.  \citet{Longuet-HigginsFoxJFM1977,McLeodStudApplMath1997} about whether the number of oscillations in the slope of a
non-limiting Stokes wave wave increases to infinity  as non-limiting Stokes wave approaches its limiting form. Note that these oscillations vanish for the limiting Stokes wave as was proven in Ref. \citet{PlotnikovTolandArchRationMechAnal2004}.



\appendix

\section{Equivalence of two form of  equation  for Stokes wave}
\label{sec:Stokeswaveequivalnce}

In this Appendix we show that both forms
\e{stokes_wave} and \e{Stokeszuzubar} of equation for Stokes wave are equivalent
to each other. Then Section \ref{sec:ClosedIntegral} implies that
equations \e{stokes_wave2} and \e{eqn:stokes22} are also equivalent
to equations  \e{stokes_wave} and \e{Stokeszuzubar}. Equation
\e{stokes_wave} was  obtained by \citet{DKSZ1996} while equation
\e{Stokeszuzubar} in slightly different forms was used by numerous
authors  including
\citet{Stokes1880sup,MalcolmGrantJFM1973LimitingStokes,SchwartzJFM1974,Longuet-HigginsFoxJFM1977,Williams1981}
and \citet{TanveerProcRoySoc1991}. Appendix
\ref{sec:restmovingframe}  explains  the derivation of equation
 \e{Stokeszuzubar} starting from basic equations
of the potential flow of ideal fluid with free surface.

 Applying the Hilbert operator $\hat H
$ \e{Hilbertdef} to equation  \e{stokes_wave} and using the relations \e{xytransform_der} and \e{yxucondition} one obtains that
\begin{equation}
\label{stokes_wavexu} c^2\tilde x_u - yx_u + \hat H[yy_u] = 0,
\end{equation}
which is equivalent to equation  \e{stokes_wave2}. We define a new variable
\begin{equation} \label{fyyudef}
f\equiv-\hat H[yy_u]
\end{equation}
and split it into two functions
\begin{equation} \label{fpm}
f=f^+ +f^-,
\end{equation}
using equations \e{ffourier}, \e{fpm0}, \e{fplus} and \e{fminus}
such that $f^+$ and $f^-$ are the functions which are analytic in
upper half-plane $\mathbb{C}^+$ and lower complex half-plane
 $\mathbb{C}^-$ of $w$, respectively.
Note that the zeroth harmonic $f_0=0$ as it follows from the definition \e{fyyudef}.
Taking the linear combination of $f$ and $\hat H f$, using equations   \e{stokes_wave}, \e{stokes_wavexu} and  \e{fyyudef}, one finds that%
\begin{equation} \label{xuHf}
x_u\hat Hf+y_u f=c^2\tilde x_uy_u.
\end{equation}
Using obvious relations

\begin{equation} \label{xuyuzu}
x_u=\frac{1}{2}(z_u+\bar z_u), \quad y_u=\frac{1}{2\I}(z_u-\bar z_u)
\end{equation}
 together with equations \e{fpm}, \e{Hfpm} one obtains from equation \e{xuHf} that
\begin{equation} \label{fzzbar}
\bar z_uf^+-z_uf^-=\frac{c^2}{4}\left (\bar {\tilde z}_u^2- {\tilde z}_u^2 \right ).
\end{equation}
Recalling that $z_u$ is analytic  in
 $\mathbb{C}^-$ and, respectively,  $\bar z_u$ is analytic and  in
 $\mathbb{C}^+$, we apply the projector \e{Projectordef} to equation   \e{fzzbar} and find that
\begin{equation}\label{fpfm1}
\begin{split}
& f^+=\frac{c^2}{4} \frac{\bar{\tilde z}_u^2}{\bar z_u},   \\
& f^-=\frac{c^2}{4} \frac{{\tilde z}_u^2}{ z_u},
\end{split}
\end{equation}
where $z_u\neq 0$ in
 $\mathbb{C}^-$ and $\bar z_u\neq 0$ in
 $\mathbb{C}^+$ because $z(w)$ is the conformal transformation in
 $\mathbb{C}^-$.

Then using equations   \e{stokes_wavexu},  \e{fyyudef}, \e{fpm},
\e{xuyuzu} and \e{fpfm1} with some algebra we recover equation
\e{Stokeszuzubar} thus completing the proof of its equivalence to
equation \e{stokes_wave}. Note that the mean-zero elevation
condition  \e{yxucondition} is essential in that proof making
equation \e{stokes_wavexu} valid. Shifting of the origin in
$y$-direction would result in the nonzero value of the mean
elevation
$y_{mean}\equiv\frac{1}{2\pi}\int\limits^{\upi}_{-\upi}\eta(x,t)\D x$.
Then one would have to replace $y$ by $y-y_{mean}$ in equation
\e{Stokeszuzubar}. E.g. \citet{TanveerProcRoySoc1991} took
$y_{mean}=-c^2/2$. A similar choice $y_{mean}=-c^2/2$ was used by
\citet{MalcolmGrantJFM1973LimitingStokes}, \citet{Williams1981} and
\citet{Plotnikov1982} up to trivial scaling of parameters.

\section{Stokes wave in the rest frame and in the moving frame}
\label{sec:restmovingframe}

Starting from \citet{Stokes1880sup}, it has been common to  write
Stokes wave equation at moving reference frame in transformed form
with a velocity potential and a stream function used as independent
variables, see e.g.
\citet{MalcolmGrantJFM1973LimitingStokes,Williams1981} and
\citet{TanveerProcRoySoc1991}. The purpose of this Appendix is to
relate that traditional form of Stokes wave equation to another form
used   by   to
\citet{DKSZ1996,ZakharovDyachenkoVasilievEuropJMechB2002}.

In physical coordinates $(x,y)$ a
velocity ${\bf v}$   of  two dimensional potential flow of inviscid
incompressible fluid is determined by a velocity potential $\Phi(x,y,t)$ as
${\bf v}= \nabla \Phi$. Here $x$ is the horizontal axis and $y$ is the vertical axis pointing upwards. The incompressibility condition $\nabla
\cdot {\bf v} = 0$ results in the Laplace equation
\begin{align} \label{laplace}
\nabla^2 \Phi = 0
\end{align}
inside fluid $-\infty<y<\eta(x,t)$. The Laplace equation is supplemented by the dynamic boundary condition (the Bernoulli equation at the free surface $y=\eta(x,t)$)
\begin{align} \label{dynamic1}
\left.\left(\dfrac{\partial \Phi}{\partial t} +
 \dfrac{1}{2}\left(\nabla \Phi\right)^2\right)\right|_{y = \eta(x,t)} + \eta = 0
\end{align}
and the kinematic boundary condition
\begin{align} \label{kinematic1}
\dfrac{\partial \eta}{\partial t}    =\left( -\dfrac{\partial \eta}{\partial x}\dfrac{\partial \Phi}{\partial x}+
\left.\dfrac{\partial \Phi}{\partial y}\right)\right|_{y = \eta(x,t)}
\end{align}
at the free surface.  In our scaled units, the acceleration due to gravity is $g=1.$ We  define the boundary value of the
velocity potential as $\left.\Phi(x,y,t)\right|_{y = \eta(x,t)} \equiv \psi(x,t)$. Equations \e{laplace},  \eqref{dynamic1} and
\eqref{kinematic1}, together with the decaying boundary condition at large depth
\begin{equation} \label{decayingphi}
\Phi(x,y,t)|_{y\to-\infty} = 0
\end{equation}
   form the closed set of equations.
Equation \e{decayingphi} implies that the rest frame is used such
that there is no average fluid flow deep inside fluid. See also Part
I \citep{DyachenkoLushnikovKorotkevichPartIStudApplMath2016} for
more details on basic equations of free surface hydrodynamics.

Consider the stationary waves moving in the positive $x$ direction (to the right) with the constant velocity $c$  so that
\begin{equation}\label{phietamoving}
\begin{split}
& \Phi=\Phi(x-ct,y),  \\
& \eta=\eta(x-ct).
\end{split}
\end{equation}

It was obtained in Ref. \citet{DKSZ1996} (see also Part I \citep{DyachenkoLushnikovKorotkevichPartIStudApplMath2016}) that
$\Psi=-c\hat Hy=c\tilde x$, where $\hat H$ is the Hilbert transform \e{Hilbertdef}.  Respectively, $\hat H\Psi=cy$. The complex
velocity potential $\Pi$ at the free surface is given by
\begin{align} \label{ComplexPotential2}
\Pi=\Psi+\I \hat H\Psi=c( x+\I y-u).
\end{align}
The analytical continuation of \eqref{ComplexPotential2} into  the lower complex half-plane $w\in\mathbb{C}^-$ is given by
\begin{align} \label{ComplexPotential3}
\Pi=c(z-w)=c\tilde z.
\end{align}

We perform a Galilean transformation to a frame moving with the velocity $c$ in the positive direction with the new horizontal coordinate $x'\equiv x-ct$ so that the velocity potential and the surface elevation turn time-independent as $\Phi(x')$ and $\eta(x')$, respectively. Alternatively, one can also define a velocity potential in the moving frame as $\tilde \Phi(x',y) =\tilde \Phi(x-ct,y)$ such that
\begin{align} \label{Phitilde}
\Phi=(x-ct)c + \tilde \Phi(x-ct,y).
\end{align}
Then equation \eqref{dynamic1} results in
\begin{align} \label{dynamic2moving}
\left.
 \dfrac{1}{2}\left(\nabla \tilde \Phi\right)^2\right|_{y = \eta(x-ct)} + \left(\eta-\frac{c^2}{2}\right ) = 0
\end{align}
and \eqref{kinematic1} gives
\begin{align} \label{kinematic2moving}
\left( -\dfrac{\partial \eta}{\partial x}\dfrac{\partial \tilde\Phi}{\partial x}+
\left.\dfrac{\partial \tilde\Phi}{\partial y}\right)\right|_{y = \eta(x-ct)}=0.
\end{align}
The decaying boundary condition \e{decayingphi} is replaced by

\begin{equation} \label{decayingphitilde}
\tilde \Phi(x',y)|_{y\to-\infty}=-c.
\end{equation}

 Equations \e{dynamic2moving}, \e{kinematic2moving} and \e{decayingphitilde} are the standard equations for Stokes
wave in the moving frame, see  e.g. \citet{MalcolmGrantJFM1973LimitingStokes,Williams1981}.
 Often small variations of equations \e{dynamic2moving}, \e{kinematic2moving} and \e{decayingphitilde}  are used such as a trivial shift of the origin in the vertical direction $\eta-\frac{c^2}{2}\to \eta,$ assuming that Stokes wave moves in negative direction (to the left)  and rescaling $c$ to one (then  the spatial period $2\upi$  is also rescaled) as was done in  \citet{MalcolmGrantJFM1973LimitingStokes}.

Similar to \eqref{Phitilde}, we define the stream function in two forms, $\Theta(x')$ and $\tilde \Theta(x')$ (in the rest frame and in the moving frame, respectively) as follows

\begin{align} \label{Thetatilde}
\Theta=c\,y+ \tilde \Theta(x-ct,y).
\end{align}
Using equations \e{Phitilde} and \e{Thetatilde} one obtains that correspondingly, that two forms of the complex velocity
potential, $\Pi(x')$ and $\tilde \Pi(x'),$  are given by
\begin{align} \label{ComplexPotential1}
\Pi=\Phi+i\Theta=cz-c^2t+\tilde \Phi+i\tilde\Theta.
\end{align}

A comparison of  \eqref{ComplexPotential3}  and \eqref{ComplexPotential1} reveals that %
\begin{align} \label{PhiThetaw}
\tilde\Pi=\tilde\Phi+i\tilde\Theta=-c(w-ct)=-cw',
\end{align}
where $w'\equiv w-ct$.
 Thus $\tilde \Pi$ is the same as $w'$  (up to the multiplication on $-c$) which explains why using the velocity potential $\tilde \Phi$ and the stream function $\tilde \Theta$ as independent variables in Refs.   \citet{Stokes1880sup,MalcolmGrantJFM1973LimitingStokes,Williams1981} is equivalent to using $w'$ as the independent variable in Ref.    \cite{DKSZ1996}. The  difference between $\Pi$ and $\tilde\Pi$ is reflected by the boundary conditions   \e{decayingphi} and \e{decayingphitilde} such that for  $\Pi$   the fluid at infinite depth has a zero velocity while for  $\tilde \Pi$ that velocity is $-c$ in the  $x$-direction. A technical advantage of working with  $\Pi$ instead of $\tilde\Pi$  in Ref.    \cite{DKSZ1996} is that the decaying boundary condition \e{decayingphi} allows to relate real and imaginary parts of $\Pi$ through the Hilbert transform for real values of $w'$ as $\Theta = \hat H \Phi $ and $\Phi =- \hat H \Theta.$ Equations  \e{dynamic2moving} and \e{PhiThetaw} results in Stokes
wave equation in the form  \e{Stokeszuzubar} after we notice that
$(\nabla \tilde \Phi)^2|_{y=\eta}=\frac{|\tilde \Pi_u|^2
}{|z_u|^2}=\frac{c^2 }{|z_u|^2}$.

\section{Tables for numerical values of $\chi_c$      for Stokes wave }
\label{sec:TablesStokesWaveschic}

\begin{table}
\begin{center}
\def~{\hphantom{0}}
\begin{tabular}{lclc}
\qquad \qquad Wave height $H/\lambda$ & Singularity position $\chi_c$ \\
0.077390566513510100664367446945009 & 0.22959283981280615879703284574991\\
0.10042675172528485854673515635249 & 0.12126855832745608069685459720991 \\
0.11396866940628458279840665192065 & 0.071654598419719678169515049620847 \\
0.12063157457100181211171486096916 & 0.050466513002046555340085106251597 \\
0.13046836752896146189584028585057 & 0.022711769117183995733113183176661 \\
0.13871124459012593791450261565795 & 0.0030056373876010407473234354599642 \\
0.14003037735536232024327827857514 & 0.0007999065189780408034349632263817 \\
0.14011096764402710691403135029555 & 0.00069951386487208337732279647662665 \\
0.14015101306439164612988663680930 & 0.00065164210434348698577048482811606 \\
0.14033404782061154512392085005894 & 0.00045087566212961727243263909506818 \\
0.14051416938624427610421738297959 & 0.00028427822364922236690177980170163 \\
0.14056584420653835444977911685203 & 0.00024252541408812956956630147113284 \\
0.14070850110629620828789822957203 & 0.00014199627497457559254018017702833 \\
0.14074703013044272779483720282718 & 0.00011868402545440790157599298606945 \\
0.14075662532618050016439516401203 & 0.0001131402886276901411780810604808 \\
0.14077748818517580368147808000934 & 0.00010145173966680681771175565637662 \\
0.14080831525231916769272562321913 & 0.000085108686515454366575393860892637 \\
0.14083140371280991872217783523764 & 0.000073606496213860270898473095913984 \\
0.14085072731982411577531399667650 & 0.000064475962982549833303295412314089 \\
0.14086825990337854565346642922133 & 0.000056590609636696915098098733019878 \\
0.14087792765270709969236336933758 & 0.00005240769363924544328892679639685 \\
0.14088586197110133631188309127224 & 0.000049063815868419517646209932713057 \\
0.14089635109209977336909824577330 & 0.00004476805660136311962064510052487 \\
0.14091001709910523062648751945506 & 0.000039388011825703454833655362993012 \\
0.14091839307555128402812965695553 & 0.000036214071851881467799287017287358 \\
0.14092032625051507844376744407087 & 0.00003549506133290811208694741093295 \\
0.14092252442341776630057428860718 & 0.0000346837089035969690548283554112 \\
0.14092514757875551525458131241416 & 0.000033724196620161039218316518297236 \\
0.14092738180637770768107092780251 & 0.000032914454078339616366407901860458 \\
0.14093056906823728426117769727974 & 0.000031771329157192593064752326105744 \\
0.14093510137194143743061264048898 & 0.0000301703287220913256069400404687 \\
0.14094119430696937198416665739014 & 0.000028063945797678144251500481216356 \\
0.14094867821783188240349944668053 & 0.000025549865907771481807832323273915 \\
0.14095352707479979419800954052129 & 0.000023964796260036642282422099761643 \\
0.14095778935504595764411825281530 & 0.000022600407539173053002286018858435 \\
0.14097009565718766875950104063752 & 0.000018816656490602043348418618380363 \\
0.14098407663748727496462567878823 & 0.00001480968355336403686583695738714 \\
0.14100153154854889551064171690484 & 0.000010273655389226364040855903301072 \\
0.14103365111671204571809985597404 & 3.5012288974834512273437793255939e-6 \\
0.14105431648358048728514606849313 & 6.0520035443913536064479745209207e-7 \\
0.14105777885488320816492860225696 & 2.9691220994639291094028846634237e-7 \\\end{tabular}
\caption{A sample of numerical values of $\chi_c$ vs. the scaled Stokes
wave height $H/\lambda.$  } \label{tab:H.chic}
\end{center}
\end{table}

Table \ref{tab:H.chic} provides a sample of the dependence of the singularity position $\chi_c$ on the scaled wave height
$H/\lambda=H/(2\pi)$ for Stokes wave. Numerical values of $\chi_c$ are obtained by the numerical procedure described in Section
\ref{sec:chicmatching}. The Pad\'e approximants from Part I \citep{DyachenkoLushnikovKorotkevichPartIStudApplMath2016} (these
approximants  are also available through the electronic attachment to Ref. \citet{DyachenkoLushnikovKorotkevichPartIArxiv2015}
and at the web link~\citet{PadePolesList}) are used for each values of $H/\lambda$. More values of  $\chi_c$  for different
values of $H/\lambda$  are also available at the web link~\citet{PadePolesList}. The accuracy of numerical values of $\chi_c$ is
at least $10^{-26}$ which is limited by the precision of  Pad\'e approximation.

We chose parameters at 1st, 3rd and 5th lines of table
\ref{tab:H.chic} to correspond Stokes waves with $c=1.03$, $1.066$
and $1.086$, respectively (here the exact values of $c$ are used).
These three particular values of parameters correspond to three
highest Stokes waves provided in table  1 of Ref.
\citet{TanveerProcRoySoc1991}. Table \ref{tab:tanveer}  reproduces
these three highest waves from table 1 of
\citet{TanveerProcRoySoc1991}, where the position of square root
branch point $\zeta=\I\chi_c$ is recovered from the parameter
$\zeta_0$ of Ref. \citet{TanveerProcRoySoc1991} as  $\chi_c = -(1 +
\zeta_0)/(1 -\zeta_0).$ Also $H$ in Ref.
\citet{TanveerProcRoySoc1991} is the half-height of Stokes wave so
it is divided by $\pi$ in table \ref{tab:tanveer}.
 The comparison of tables \ref{tab:H.chic} and  \ref{tab:tanveer}   reveals that while all digits except the last one or two agree for two smaller values of $H/\lambda,$   but the agreement looses one more digit with the increase of $H/\lambda$. It is possible that Ref. Tanveer (1991) expected that loss of numerical precision because the number of digits provided  in table 1 of Tanveer (1991)  decreases with the increase of $H/\lambda$.

Table
\ref{tab:asheet} provides a sample of numerical values of
$a_{-,2n,0}$ and  $a_{+,2n+1,0}$ for four different values of
$\chi_c$ corresponding to  table \ref{tab:H.chic}. These numerical
values of $\chi_c$ are obtained by the numerical procedure described
in Section   \ref{sec:apmmatching}. For brevity only 16 digits of
the numerical precision are shown.

\begin{table}
  \begin{center}
\def~{\hphantom{0}}
  \begin{tabular}{ccc}
         $H/\lambda$   &   $c$ & $\chi_c$ \\
0.07739055          &   1.0300 &        0.22958 \\
0.1139758          &   1.0660 & 0.071667\\
0.13055          &   1.0860 & 0.022769\\
        \end{tabular}
  \caption{Parameters of three highest Stokes waves of Table  1 from \citet{TanveerProcRoySoc1991}. Units are converted to the notation of this paper with the same number of digits kept as in  Ref. \citet{TanveerProcRoySoc1991}. }
  \label{tab:tanveer}
  \end{center}
\end{table}

\begin{table}
  \begin{center}
\def~{\hphantom{0}}
  \begin{tabular}{lcccc}
      ~  & $\chi_c=0.12126\ldots$   &   $\chi_c=0.05046\ldots$ & $\chi_c= 0.000242\ldots$& $\chi_c= 2.969\ldots \times 10^{-7}$\\[3pt]
       $a_{-,2,0}$ & 1.947517181530394 & 1.332875450393561 & 0.616114648091185 & 0.5967616372529635 \\
       $a_{+,3,0}$   &  1.933089192507101 & 1.395722669719572 & 0.6227276830074182 & 0.5968472222666076 \\
       $a_{-,4,0}$  &   2.823744469669705 & 1.830765178354924 & 0.630550992725188 & 0.5969268580437934 \\
       $a_{+,5,0}$   &  2.715883020102187 & 1.841263995239744 & 0.6356646908933044 & 0.5969952862738779 \\
       $a_{-,6,0}$ &    3.541346294820654 & 2.238582675537623 & 0.642377214720984 & 0.5970622067844041 \\
  \end{tabular}
  \caption{A sample of numerical values of $a_{-,2n,0}$ and  $a_{+,2n+1,0}$, $n=1,2,3$, for different  $\chi_c$. More accurate numerical values of $\chi_c$ can be recovered from table \ref{tab:H.chic}.  }
  \label{tab:asheet}
  \end{center}
\end{table}


\end{document}